\documentclass[10pt,letterpaper]{article}
\usepackage[top=0.85in,left=2.75in,footskip=0.75in]{geometry}

\usepackage{amsmath,amssymb}

\usepackage{changepage}

\usepackage[utf8x]{inputenc}

\usepackage{textcomp,marvosym}

\usepackage{cite}

\usepackage{nameref,hyperref}

\usepackage[right]{lineno}

\usepackage{microtype}
\DisableLigatures[f]{encoding = *, family = * }

\usepackage[table]{xcolor}

\usepackage{array}

\newcolumntype{+}{!{\vrule width 2pt}}

\newlength\savedwidth

\raggedright
\usepackage[aboveskip=1pt,labelfont=bf,labelsep=period,justification=raggedright,singlelinecheck=off]{caption}

\makeatletter
\renewcommand{\@biblabel}[1]{\quad#1.}
\makeatother

\usepackage{lastpage,fancyhdr,graphicx}
\usepackage{epstopdf}
\fancyheadoffset[L]{2.25in}
\fancyfootoffset[L]{2.25in}
\usepackage{bm}

\def\be{\begin{equation}}
\def\ee{\end{equation}}
\def\bea{\begin{eqnarray}}
\def\eea{\end{eqnarray}}
\def\bma{\begin{mathletters}}
\def\ema{\end{mathletters}}

\def\q0{\underline{0}}

\def\R{\mathbb{R}}

\def\one{\leavevmode\hbox{\small1\normalsize\kern-.33em1}}

\definecolor{nred}{rgb}{0.9,0.1,0.1}

\definecolor{blue}{rgb}{0.0,0.0,0.9}

\begin{document}
\vspace*{0.2in}
\begin{flushleft}
{\Large
\textbf\newline{Disease control as an optimization problem} 
}
\newline
\\
Miguel Navascu\'es\textsuperscript{1*},
Costantino Budroni\textsuperscript{2,1},
Yelena Guryanova\textsuperscript{1}
\\
\bigskip
\textbf{1} Institute for Quantum Optics and Quantum Information (IQOQI), Austrian Academy of Sciences, Boltzmanngasse 3, 1090 Vienna, Austria
\\
\textbf{2} Faculty of Physics, University of Vienna, Boltzmanngasse 5, 1090 Vienna, Austria
\bigskip

%
%
* miguel.navascues@oeaw.ac.at

\end{flushleft}

\section*{Abstract}

In the context of epidemiology, policies for disease control are often devised through a mixture of intuition and brute-force, whereby the set of logically conceivable policies is narrowed down to a small family described by a few parameters, following which linearization or grid search is used to identify the optimal policy within the set. This scheme runs the risk of leaving out more complex (and perhaps counter-intuitive) policies for disease control that could tackle the disease more efficiently. In this article, we use techniques from convex optimization theory and machine learning to conduct optimizations over disease policies described by hundreds of parameters. In contrast to past approaches for policy optimization based on control theory, our framework can deal with arbitrary uncertainties on the initial conditions and model parameters controlling the spread of the disease, and stochastic models. In addition, our methods allow for optimization over policies which remain constant over weekly periods, specified by either continuous or discrete (e.g.: lockdown on/off) government measures. We illustrate our approach by minimizing the total time required to eradicate COVID-19 within the Susceptible-Exposed-Infected-Recovered (SEIR) model proposed by Kissler \emph{et al.} (March, 2020).

\section*{Authors' summary}
We consider the problem of finding the optimal government policy to fight a contagious disease. To this end, we develop a number of fully general computational tools, based on notions from machine learning and convex optimization. Given a mathematical model for the disease spread, the government interventions available (e.g., lockdowns or vaccination campaigns), a cost function (e.g., the time until disease extinction, the total death toll) and a list of constraints (e.g.: the healthcare system not collapsing throughout the whole course of the epidemic), our algorithms identify the sequence of government interventions that minimizes the cost function, while respecting all considered constraints. Notably, our algorithms can integrate uncertainties in the model parameters and stochastic models of disease spreading; this separates our work from past proposals and makes our methods suitable for practical use. We observe that optimal policies tend to be extremely complex and suggest that in the future the process of policy generation could be partly automated. 

\section{Introduction}
The COVID-19 pandemic has already caused over four million deaths worldwide. The effects of the virus have been widespread and substantial, from the collapse of healthcare systems \cite{un_framework,yemen,salvador} to the enforcement of isolation and quarantine. In the case of Nepal, the national lockdown lasted for $120$ days uninterrupted \cite{kathmandu}. 

In these circumstances, identifying reliable and effective disease control policies is of utmost importance. Here by ``policy'' we mean a deliberate intervention intended to mitigate the effects of a disease as it runs its course. In much of the mathematical literature on epidemiology, the process of generating a policy is as follows \cite{book}: (1) based on expert intuition, a number of suitable policies to control the disease are proposed; (2) the impact on the population of each of the considered policies is assessed through dynamical models of disease spread; (3) the outcomes of all policies are compared and a decision is taken as to which one is deemed to be the best.

The main advantage of this three-step process is that the final recommended policy is comprehensible, i.e., it can be interpreted and explained. Moreover, for simple on/off policies, one can sometimes derive analytic results \cite{analytic_discrete}. The process has, nonetheless, two disadvantages. First of all, the class of policies devised by an human could well be suboptimal, since the optimal policy (under some figure of merit) could be extremely complicated and counter-intuitive. Second, the method generally requires one to numerically simulate each policy, and so is inapplicable when the considered class of policies depends on many control parameters: due to the exponentially large number of conceivable policies, by the time one finds the optimal disease control policy, it would be too late to enforce it.

Other approaches for disease control rely on optimal control theory to identify a suitable policy (see, e.g., \cite{control_theory1, control_theory2,control_theory3,control_theory4}). The starting point of all these works is that both the initial conditions (namely, the number of individuals infected, exposed, etc.) and the model parameters (such as the disease's reproduction number) specifying the spread of the disease are known with high precision. This requirement is never met in a real-life epidemic, especially close to the outbreak, when the uncertainty in the disease's reproduction number can be very high \cite{R_0,R_0_bis}. Policies derived through optimal control theory are thus not guaranteed to have the desired effects in practice.

An additional disadvantage of optimal control theory is that government measures in the model cannot be constrained to be discrete. It is true that some optimal control problems admit a discrete solution, a so-called bang-bang control, but one cannot enforce this property on the solution \emph{a priori}. This makes optimal control theory all the more impractical, for discrete measures, such as a lockdown that is either on or off, have so far dominated global efforts to control the COVID-19 epidemic \cite{lockdown}. Last but not least, optimal control theory can only handle scenarios where the policies  vary over time continuously, thus not making it compatible with observed government measures to control COVID-19, which for the most part have been applied on a discrete, weekly basis. 

In this paper we introduce a general framework that maps any disease control scenario to an optimization problem. Contrary to the optimal control approach, our framework can accommodate constraints on the disease dynamics which must hold for whole regions of the initial conditions and the disease's model parameters. Our framework can enforce policies to be weekly and/or discrete. Invoking tools from optimization theory and machine learning \cite{deep_learning}, we propose efficient heuristics to solve the optimization problem and hence identify the government policy that best controls the disease. 

To illustrate the power of our approach, we use these optimization techniques to generate long-term plans to fight COVID-19, under the assumption that the disease's dynamics are accurately captured by a variant of the Susceptible-Exposed-Infected-Recovered (SEIR) compartmental model~\cite{book} proposed in~\cite{social_distancing}. Our results confirm that optimal policies tend to be too complicated to be devised by a human. 

In reality, most epidemiological models only provide short-term approximations to the spread of the disease, with long term projections becoming less and less reliable~\cite{predictability}. In addition, notwithstanding the enormous knowledge gathered since the initial COVID-19 outbreak, many questions remain to be answered regarding the correctness and accuracy of compartmental models such as SEIR: their basic assumptions (e.g., are recovered patients temporarily or permanently immune to the disease?); the actual value of the model's parameters (e.g., the basic reproduction number $R_0$); and the role of variables not modeled (e.g., age, geographic distribution, contact tracing policies, role of superspreaders). 

In this regard, the goal of this work is not to propose a concrete government policy, but rather to present an efficient method to obtain an optimal one, given all the available information. To estimate the effect of our methods in a realistic scenario, we conduct a numerical simulation where we re-calculate the optimal policy plans every month, based on new, incoming data. The very final policy plan that we present requires less stringent physical distancing measures compared with the one which is not re-calculated every month. All the code used in our simulations is freely available at~\cite{githuburl}.

\section{The framework in a nutshell}\label{sec:frame}

Our starting point is an epidemic that affects a closed population. This assumption is not limiting, since a large ensemble of population centres where individuals are free to commute can also be modeled as a closed system \cite{open_systems}. To gain an understanding of how the disease spreads, it is standard to divide the population into different sectors or compartments \cite{book} (see Fig. \ref{fig:schematic}). One can define, e.g., the compartment of all those individuals who are currently infected. This compartment can, in turn, be sub-divided into different compartments, such as symptomatic/asymptomatic. Once the number of relevant compartments is fixed, one can estimate the occupation of each of them and arrange the resulting numbers in a vector $\bm{x}$. The disease is subsequently analysed by looking at how $\bm{x}$ changes with time.

In order to control or even extinguish an epidemic, governments can enforce a number of different measures: mass vaccination, physical (or social \cite{WHO_Press2020}) distancing measures, or even a full lockdown, are common examples of interventions aimed to fight the disease. When and to which degree such measures are applied is determined by the \emph{disease control policy}. Consider a disease control policy based on vaccination campaigns, where the intervention consists of vaccinating a number of individuals per day, across all compartments, i.e., without distinguishing between susceptible, exposed, recovered, and so on. Let $v(t)$ be the fraction of the total population vaccinated on day $t$. This function, between the initial and final times $t_0$ and $t_f$ (i.e. on the interval $[t_0, t_f]$), determines the government's vaccination policy. Similarly, let $s(t)$ take the value $1$ if the country is in lockdown on day $t$ and $0$ if it is not. Then the government's lockdown policy between times $t_0$ and $t_f$ corresponds to the function $s(t)$.
\begin{align}\label{eq:s}
s(t) = 
\begin{cases}
0 & \text{lockdown off } \\
1 & \text{lockdown on} 
\end{cases} 
\qquad \;
 t \in [t_0, t_f]
 \,.
\end{align}
If the government is intervening both through vaccination and lockdown, then its disease control policy will be identified by \emph{both} functions $v(t)$ and $s(t)$. Note that, whereas $v(t)$ can take a continuum of values, $s(t)$ can only have finitely many. In the following, we will call the first class of interventions \emph{continuous}; and the second, \emph{discrete}. In general, a policy for disease control will combine both kinds of measures, but, as we will see, optimizing over one class or the other requires very different techniques.

The above are instances of \emph{non-adaptive policies} for disease control, because the functions $s, v$ only depend on the time $t$, and not, e.g., on the current death rate. A general \textit{adaptive} policy for disease control would take into account the whole past history of data gathered by the government before deciding what to do at each step. Although in the following all our proposed policies are non-adaptive, the formalism we introduce allows one to optimize over adaptive policies as well.

In conclusion, a disease control policy can always be associated to a time-dependent vector function $\bm{\alpha}$ and perhaps some other observed variables $\bm{o}$, where each vector entry represents a type of government intervention at time $t$. In turn, we can use a variable vector $\bm{\mu}\in \R^n$ to parametrize the class of considered policies, that is, $\bm{\alpha}(t, \bm{o})=\bm{\alpha}(t, \bm{o};\bm{\mu})$. Since $\bm{\mu}$ completely determines the policy $\bm{\alpha}$, we can also regard the parameters $\bm{\mu}$ as the disease policy. We will do so from now on. 

The applied policy $\bm{\mu}$ is assumed to influence the compartment occupation within the time interval $[t_0, t_f]$. That is, $\bm{x}$ is both a function of $t$ and $\bm{\mu}$. In this work we are interested in devising policies for disease control which guarantee that the spread of the disease evolves under certain conditions. For example, any country has a fixed number of critical care beds, which we denote by ${\cal B}_c$. At each time $t\in [t_0, t_f]$, it is desirable that the number of individuals admitted to critical care in hospitals, ${\cal C}(t)$, does not exceed that capacity. That is, we require that

\be
{\cal C}(t)\leq {\cal B}_c, \quad\mbox{ for } t\in [t_0, t_f].
\label{cc_constr}
\ee
\noindent We will call any such condition on the policy, or on its effect on the evolution of the disease, a \emph{constraint}. Further examples of relevant constraints are the requirement that the number of planned vaccinations does not exceed the government's total supply, or that the death rate does not surpass a given threshold.

Finally, among all policies satisfying the desired constraints, we typically wish to identify the one that minimizes a certain quantity. For example, for many diseases, a simple physical distancing policy that satisfies the constraint (\ref{cc_constr}) consists of declaring a lockdown throughout the whole time interval $[t_0, t_f]$, i.e., $s(t)=1$ for $t\in [t_0, t_f]$. This policy is arguably impractical, difficult to enforce and harmful to its citizens' psychological health, as well as to the national economy. More rationally, one is interested in finding alternative disease policies which, while respecting the critical care occupation constraint, minimize the number of days of lockdown. Alternatively, one may wish to minimize the total number of deaths during the interval $[t_0, t_f]$, or the number of infected people. In general, the figure of merit or quantity that we wish to minimize will be a complicated functional of the considered policy $\bm{\alpha}$ and $\bm{x}$. We will call this functional the \emph{objective function}. A sufficiently complex objective function, together with the appropriate optimization constraints, can meet any conceivable set of societal demands.

The optimization problem sketched above is mathematically ill-defined, unless we specify how $\bm{x}$ varies with the parameters $\bm{\mu}$ determining the policy. In order to predict the natural course of a disease or how a given policy might affect its spread, 
experts make use of mathematical models. In this paper, we will mainly be concerned with \emph{deterministic compartmental models}, but we will show also how our method extends to stochastic models. In these models, the whole population is divided into a number of basic compartments and the interactions between those compartments are modeled, in the deterministic case, through a system of ordinary differential equations. Given the occupation $\bm{x_0}$ of the compartments at time $t_0$, these models allow us to compute the value of $\bm{x}$ at any instant $t\in[t_0, t_f]$ as a function of the policy $\bm{\mu}$. That is, each model provides an implicit functional relation of the form
\begin{align}\label{eq:compartments}
\bm{x}=\bm{x}(t; \bm{\mu}, \bm{x_0}).
\end{align}

Past literature on disease control has made extensive use of compartmental models to recommend specific strategies for policy-makers in an effort to control the spread of a disease. In many cases, suitable policies are devised through a mixture of intuition and grid search, see \cite{vaccination_SEIR, pulse_vaccination, foot_and_mouth, social_distancing}. The starting point is a family of disease control policies with one or two unknowns. For instance, in pulse vaccination \cite{pulse_vaccination}, those unknowns are the time intervals between two mass vaccinations and the vaccination rate. In this paper, we propose a scheme that allows for the optimization over policies specified by thousands of parameters in the space of a few hours. Our scheme, detailed in the following sections, is based on a standard tool in optimization theory and machine learning known as \emph{gradient descent} \cite{subgradient, stoch}. Starting with a rough guess for the optimal policy $\bm{\mu}^{(0)}$, gradient descent methods generate a sequence of policies $\bm{\mu}^{(1)}, \bm{\mu}^{(2)},...$ which typically exhibit increasingly better performance. Although the gradient method is not guaranteed to converge to the optimal policy, after many iterations it generates solutions that are good enough for many practical purposes. In fact, gradient descent is the method most commonly used to train deep neural networks \cite{deep_learning} and support vector machines \cite{support}. 

Importantly, the computational resources required to carry out the gradient descent method are comparable to the cost of running a full simulation between times $t_0$ and $t_f$ with the considered disease model. Furthermore, the necessary computations can be \emph{parallelized} for policies depending on many parameters. Although the focus of this paper is on compartmental disease models, our main ideas can also be used to understand ecological systems undergoing more complex dynamics \cite{raccoons}, see Appendix \ref{spatial_app}. Even in such complicated scenarios, the former scaling laws hold: provided that we can run the considered disease model, we can apply the gradient method to optimize over policies of disease control.

\section{The gradient method}
\label{gradient_sec}
We next introduce the gradient method and show how one can use it to tackle an abstract optimization problem. Given functions $f, g_i:\R^n\to\R$, for $i=1,...,K$, consider the following task:
\begin{align}
&\min_{\bm{\mu}} f(\bm{\mu})\nonumber\\
\mbox{such that } &g_i(\bm{\mu})\leq h_i,\mbox{ for } i=1,\ldots,K.
\label{optim_problem}
\end{align}
Each of the conditions $g_i(\bm{\mu})\leq h_i$ is called a \emph{constraint}.

Define $M=\{\bm{\mu}: g_i(\bm{\mu})\leq h_i, i =1,...,K\}$. If $f$ is sub-differentiable, a simple heuristic to solve this problem is the projected gradient method \cite{subgradient}. Call $\bm{\mu}^\star$ the solution of the problem. Starting from an initial guess $\bm{\mu}^{(0)}$, the gradient method generates a sequence of values $(\bm{\mu}^{(k)})_k$ with the property that $\lim_{k\to\infty}\bm{\mu}^{(k)}=\bm{\mu}^\star$, provided that $M, f$ are, respectively, a convex set and a convex function \cite{subgradient}. The sequence $(\bm{\mu}^{(k)})_k$ is generated recursively via the relation
\be
\bm{\mu}^{(k)} =\pi_M\left(\bm{\mu}^{(k-1)} - \epsilon \nabla_{\bm{\mu}} f(\bm{\mu}^{(k-1)})\right),
\label{gradient}
\ee
where, for any set $A\in\mathbb{R}^n$, $\pi_A(\bm{z})$ denotes the point $\bm{y}\in A$ that minimizes the Euclidean distance, i.e., $\min_{\bm{y}\in A}\|\bm{y}-\bm{x}\|_2$. 

Unfortunately, in the problems we encounter in this paper, $f$ is not convex and, sometimes, neither is $M$. This implies that the sequence output by the projected gradient method is not guaranteed to converge to the solution of the problem, but to a local minimum thereof.

In machine learning, optimization problems with non-convex objective function $f$ and $M=\R^n$ are legion. To solve them, deep learning practitioners typically use variants of the gradient method sketched above. One of these variants, Adam \cite{adam}, is extensively used to train neural networks. 

Adam works as follows. Starting with the null vectors $\bm{m^{(0)}}, \bm{v^{(0)}}\in \R^n$, vector sequences are generated according to the following iteration rule:
\begin{align}
&\bm{G}^{(k)}=\nabla_{\bm{\mu}} f(\bm{\mu}^{(k-1)})\nonumber\\
&\bm{m}^{(k)} = b_1\bm{m}^{(k-1)}+(1-b_1)\bm{G}^{(k)}\nonumber\\
&\bm{v}^{(k)} = b_2\bm{v}^{(k-1)}+(1-b_2)\bm{G}^{(k)}\odot \bm{G}^{(k)}\nonumber\\
&\hat{\bm{m}}^{(k)}=\frac{\bm{m}^{(k)}}{1-b_1^k}\nonumber\\
&\hat{\bm{v}}^{(k)}=\frac{\bm{v}^{(k)}}{1-b_2^k}\nonumber\\
&\bm{\mu}^{(k)}=\bm{\mu}^{(k-1)}-\epsilon\frac{\hat{\bm{m}}^{(k)}}{\sqrt{\hat{\bm{v}}^{(k)}}+\delta},
\label{adam}
\end{align}
where $G^{(k)}\odot G^{(k)}$ denotes the vector of the element-wise product; similarly, the fraction and square root in the definition of $\bm{\mu}^{(k)}$ are defined element-wise. Recommended values for the free parameters $\epsilon, b_1,b_2,\delta$ are $\epsilon =0.001$, $b_1=0.9$, $b_2=0.999$ and $\delta = 10^{-8}$ \cite{adam}.

Like the projected gradient method, Adam is not guaranteed to converge to the optimal solution of the problem. However, provided that the initial conditions and the learning rate $\epsilon$ are chosen with care, Adam has been observed to typically output a local minimum that is ``good enough''.

Note that, by taking $M=\R^n$, it is not clear how to enforce that the solution satisfies constraints of the form $g_i(\bm{\mu})\leq h_i$. The answer is to include those constraints as penalties in the objective function. That is, rather than minimizing $f$, we apply Adam to minimize the function
\be
f(\bm{\mu}) + \sum_{i=1}^K \nu_i(g_i(\bm{\mu})-h_i)^+,
\ee
where $\nu_i\gg 1$ and $z^+$ denotes the positive part of $z$, i.e., $z^+=z$ for $z> 0$; otherwise, $z^+=0$. For high enough values of $\{\nu_i\}_i$, the solution of the problem will just violate the constraints slightly, i.e., $h_i-g(\bm{\mu}^\star)\in [-\delta, \infty)$, for $\delta\ll 1$. If no violation whatsoever is desired then one can instead optimize over a function of the form

\be
f(\bm{\mu}) + \sum_{i=1}^K \nu_i(g_i(\bm{\mu})-h_i+\delta')^+,
\ee
\noindent with $\delta'>0$.

In some situations, the objective function $f$ will be complicated to the point that computing its exact gradient is an intractable problem. It might be possible, though, to generate a random vector $\tilde{\nabla}_{\bm{\mu}}f\in\R^n$ with the property
\be
\left\langle \tilde{\nabla}_{\bm{\mu}}f\right\rangle = \nabla_{\bm{\mu}}f.
\ee

In such a predicament, we can solve the original optimization problem (\ref{optim_problem}) through stochastic gradient descent methods \cite{stoch}. Stochastic gradient descent consists of applying the considered gradient method, with the difference that, every time that the method requires the gradient of $f$, we input the random variable $\tilde{\nabla}_{\bm{\mu}}f\in\R^n$ instead. Namely, it suffices to replace $\nabla_{\bm{\mu}} f(\bm{\mu}^{(k-1)})$ by $\tilde{\nabla}_{\bm{\mu}} f(\bm{\mu}^{(k-1)})$ in the iterative equations (\ref{gradient}), (\ref{adam}). As before, if both $M$ and $f$ are convex, stochastic gradient descent methods are guaranteed to converge to the optimal solution of problem (\ref{optim_problem}) \cite{stoch}.

\section{Overview of techniques}
\label{overview}
Our novel contribution is to show how to apply the gradient method to any disease control scenario and optimize over \textit{discrete} or \textit{continuous} response policies. In both cases the task is first to identify a suitable functional $A$, that is to be minimised, and then to show how to compute the gradient of it, which appears in the first line of the Adam algorithm in Eq.~\eqref{adam}, such that the algorithm can step to the next iteration. 

In a discrete policy, the parameters describing the government intervention can only take a finite number of values. Eq.~\eqref{eq:s} is an example of a discrete policy, because on day $t$ the government can either declare a lockdown ($s(t)=1$) or declare no lockdown ($s(t)=0$): we allow for no values of $s(t)$ inbetween. 

On the other hand, a continuous policy is one in which the parameters that define the policy, which in general we denote by the vector $\bm{\mu}$, are allowed to vary over a continuum. For example, the fraction $v(t)$ of the population vaccinated on day $t$ can take any value between $0$ and $1$.

Regardless of whether we chose to optimise over discrete or continuous policies, our starting point is an ordinary differential equation of the form
\be
\frac{dx^i}{dt} = G^i(t,\bm{x};\bm{\mu});
\quad i=1,...,m
\,,
\label{eq:diffeq}
\ee
where the entries of the vector $\bm{x}$ represent the occupations of the different compartments of a disease model
and $\bm{\mu}$ represents a continuous or discrete parametrization of the effects of a given policy. Let $\bm{x}(t;\bar{\bm{\mu}},\bm{x}_0)$ be the solution of Eq.~(\ref{eq:diffeq}) with initial conditions $\bm{x}(0)=\bm{x}_0$ and $\bm{\mu}=\bar{\bm{\mu}}$, where the initial policy condition is typically taken to be `lockdown off'. 
We consider the problem of finding the parameters $\bm{\mu}^\star$
such that $\bm{x}(t;\bm{\mu}^\star,\bm{x}_0)$ minimizes a given functional $A$. This functional defines how we wish to control the disease and what for: it might represent the number and duration of lockdown periods, etc. If the initial conditions $\bm{x}_0$ are known, then it makes sense to consider functionals of the form
\be
A(\bm{\mu},\bm{x}_0) =\int_{t_0}^{t_f} dt
\;
{\cal L}(t, \bm{\mu}, \bm{x}(t;\bm{\mu},\bm{x}_0)) 
+ \hat{A}(\bm{\mu})
\,,
\label{functional1cont}
\ee
%
where, as discussed in Section~\ref{gradient_sec}, constraints which we wish the solution $\bm{x}(t;\bm{\mu}^\star,\bm{x}_0)$ to satisfy are incorporated into ${\cal L}(t, \bm{\mu}, \bm{x}(t;\bm{\mu},\bm{x}_0))$. For instance, if we want that the number of patients in critical care does not exceed the number of beds, i.e. Eq.~\eqref{cc_constr} to hold, then one of the terms in ${\cal L}$ would be
\be
\rho(t) \left(C(t;\bm{\mu}, \bm{x}_0)- B_c\right)^+,
\label{rho_C_c}
\ee
with $\rho(t)\gg 1$. Similarly, to model a constraint on the vaccine supply of the form $\int dt \; v(t;\bm{\mu})\leq V$, we would add the term
\be
\lambda \left(\int dt \;
v(t;\bm{\mu})- V\right)^+,
\label{supply}
\ee
\noindent with $\lambda\gg 1$.

As it turns out, minimizing functional (\ref{functional1cont}) (or more complicated ones, see Appendix \ref{optimization_app}) requires different techniques depending on whether the desired policy is continuous or discrete.

\subsection{Continuous policies}
If all the parameters $\bm{\mu}$ defining the policy are allowed to take arbitrary values in $\R^n$ and their effect on the disease's equations of motion (\ref{original}) is differentiable, then one can simply apply gradient descent methods to minimize $A(\bm{\mu},\bm{x}_0)$. The only difficulty stems in computing $\nabla_{\bm{\mu}} A$ (i.e., the first line of the Adam algorithm in Eq.~\eqref{adam}). For functionals of the form (\ref{functional1cont}), we have
\be
\frac{\partial A}{\partial \mu_j}=\int_{t_0}^{t_f} dt \left(\frac{\partial {\cal L}(t, \bm{\mu}, \bm{x}(t;\bm{\mu}, \bm{x}_0))}{\partial \mu_j}+ \sum_i \frac{\partial {\cal L}(t, \bm{\mu}, \bm{x}(t;\bm{\mu}, \bm{x}_0))}{\partial x^i}\frac{\partial x^i}{\partial \mu_j}\right)+\frac{\partial \hat{A}(\bm{\mu})}{\partial\mu_j}.
\label{eq:partial}
\ee
In order to evaluate this quantity to step the algorithm, the challenge is to compute the derivatives $\frac{\partial x^i}{\partial \mu_j}$. In Appendix~\ref{optimization_app} we show that these derivatives arise as the solution of a system of ordinary differential equations, of complexity comparable to that of Eq. (\ref{eq:diffeq}). We also explain how to use stochastic gradient descent to deal with scenarios where the model parameters (e.g., the basic reproduction number $R_0$) and/or the initial conditions $\bm{x}_0$ are only known to lie within some bounds, or when the evolution is stochastic. In either case, once a method to compute \eqref{eq:partial} is established (or, in the case of uncertainty in the initial conditions, some unbiased statistical estimator thereof (see Appendix \ref{optimization_app})), one can simply run the vanilla gradient descent (\ref{gradient}) or the Adam algorithm (\ref{adam}) to arrive at a quasi-optimal policy.\\

\subsection{Discrete policies}
When some or all the entries of $\bm{\mu}$ are restricted to take values on a finite, fixed set, it is clear that one cannot apply gradient descent methods straightforwardly in order to minimize the objective function. In fact, optimizations over discrete variables are, in general, a very difficult endeavor: it can be argued that problems which appear simple do not have an efficient solution \cite{SAT}. Nonetheless, in Appendix~\ref{discrete_app} we present two heuristics to tackle such optimizations in the context of policies for disease control. 

The first heuristic consists of mapping the original minimization problem to an optimization over \emph{probabilistic} policies, whereby the government decides how to intervene by sampling a discrete probability distribution, which is continuously dependent on a number of auxiliary (continuous) parameters. By applying stochastic gradient descent methods over such auxiliary parameters, we can find the probabilistic policy that minimizes the average value of the objective function. As shown in Appendix~\ref{discrete_app}, independently of the initial conditions, the gradient method is guaranteed to converge to a deterministic policy.

The second heuristic is tailor-made for optimizations over lockdown (discrete) policies, although it can be easily extended to deal with arbitrary discrete policies. Consider a scenario in which the government is allowed to declare or lift a lockdown at any time. In this case, we can parametrize the policy by the duration of each lockdown phase and the pauses in between, and use gradient descent methods to arrive at the optimal times. In Appendix~\ref{discrete_app} we show that estimating the correspondent gradients only requires a minor modification of the methods developed for continuous policies.

\section{Application: COVID-19}
\label{applications}
To illustrate the use of gradient descent for policy design, we consider a variant of the extended ``susceptible-exposed-infected-recovered'' (SEIR) disease model proposed in~\cite{social_distancing} to predict the impact of COVID-19 in the USA, a schematic of which is shown in Fig.~\eqref{fig:schematic} and the details of which appear in Appendix \ref{models_app}. Each state variable $X$ in the diagram corresponds to the \emph{fraction} of the whole population ${\cal N}_{pop}$ in the corresponding diagram. From now on, we denote intensive quantities like $X$ with a normal mathematical font, while extensive quantities ${\cal X}=X\times {\cal N}_{pop}$ are to be represented with calligraphic letters. The clinical parameters of the model in Figure \ref{fig:schematic}, such as recovery and hospitalization rates, were estimated in \cite{medical_parameters, KisslerScience2020}, based on early reports from COVID-19 cases in the UK, China and Italy. Following the model in~\cite{social_distancing}, the disease transmission is assumed to be seasonal by analogy with the known behavior of betacoronaviruses such as HCoV-OC43~\cite{KisslerScience2020}, with a baseline reproduction number between $2$ and $2.5$, following fits of the early growth-rate of the epidemic in Wuhan \cite{R_0,R_0_bis}. 

\begin{figure}[h]
  \centering
  \includegraphics[width=14 cm]{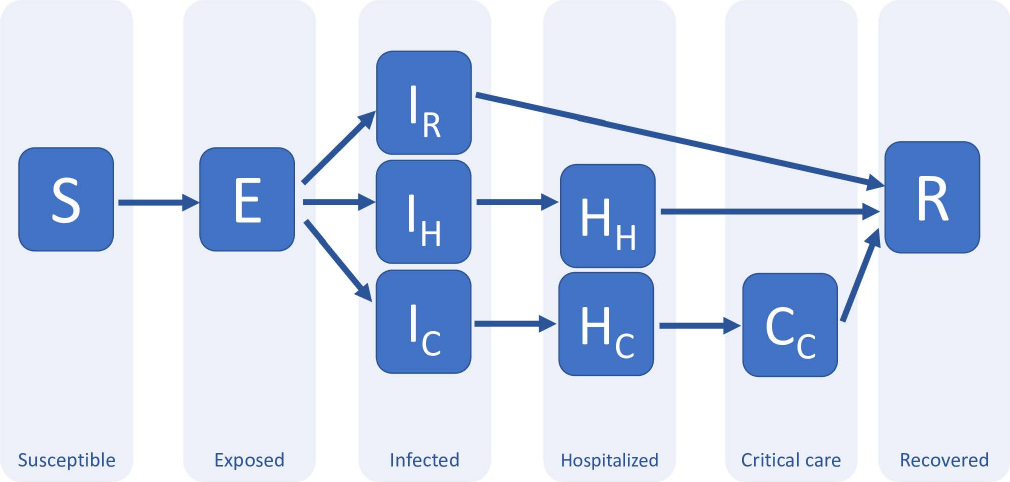}
  \caption{\textbf{A possible compartment model for COVID-19 (adapted from \cite{social_distancing})}. The main compartments are: susceptible; exposed; infected; hospitalized; critical and recovered. This compartmental splitting captures different possible evolutions as well as time delays between transitions. The ``infected'' compartment, for instance, contains those who will recover without hospitalization ($I_R$); those who will be hospitalized but won't need critical care ($I_H$); and those who will end up receiving critical care ($I_C$). The ``exposed'' compartment is introduced here to model the time delay between the exposure to the disease and the development of symptoms (incubation period), in particular, the possibility of infecting others, which is what is relevant for the model. In this model, the compartment ``recovered'' includes both dead and alive individuals; in principle, it could be sub-divided further.}. 
  \label{fig:schematic}
\end{figure}

A relevant compartment in this model is ${\cal C}(t)\equiv C_C(t)\times {\cal N}_{pop}$, the population occupying a critical care bed at time $t$. The patients sent to critical care cannot breathe unassisted, and thus it is fundamental to ensure that such capacity is not surpassed, namely, that the constraint (\ref{cc_constr}) holds. For our simulations, we chose a population size of ${\cal N}_{pop}=47$ million and ${\cal B}_c=9.5\times 10^{-5}\times {\cal N}_{pop}$. That is, we assumed that the healthcare system provides $95$ critical care beds per million inhabitants. This is a good approximation to the healthcare capacity of many European countries, as well as the USA.

According to the chosen model, without intervention, the number of citizens requiring a bed in a critical care unit evolves according to Figure \ref{cc_nat_fig} (see Appendix \ref{models_app} for the exact initial conditions of our numerical simulations). As the reader can appreciate, between the third and seventh months, the number of people in need of critical care exceeds the capacity of the considered healthcare system by $18$ times. 

\begin{figure}
  \centering
  \includegraphics[width=14 cm]{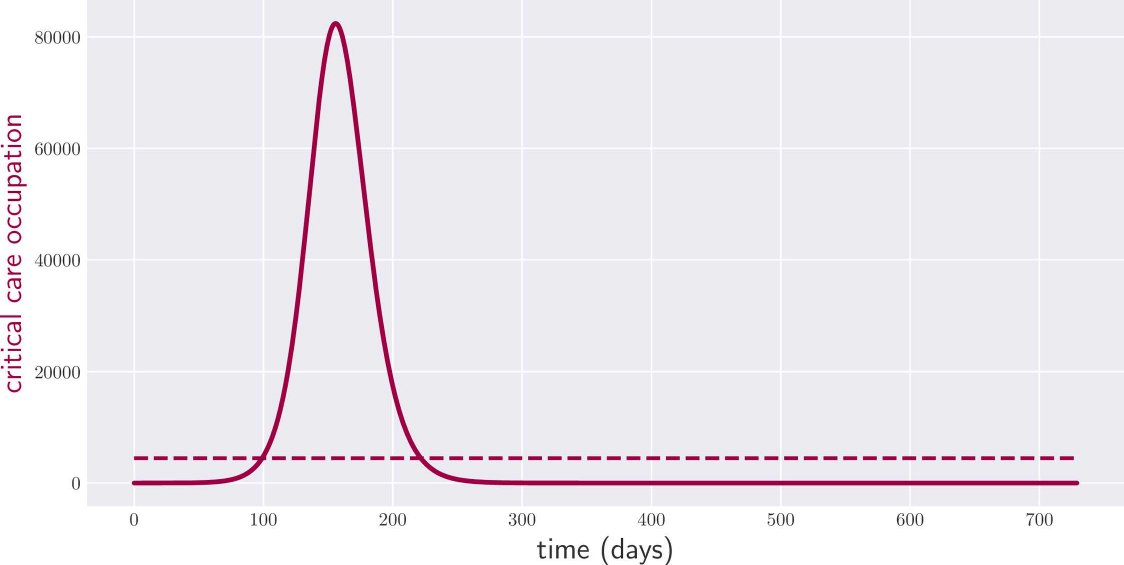}
  \caption{\textbf{Occupation of critical care beds over two years with no policy intervention.} The red dashed line indicates the critical care capacity of the healthcare system.}
  \label{cc_nat_fig}
\end{figure}

Vaccines against COVID-19 were not available during the first year of the pandemic. Hence initially most governments opted to control the disease via distancing measures and/or lockdowns. The effect of implementing a policy $s(t)$
is to multiply the disease's basic reproduction number $R_0$ by a factor of $r$ \cite{social_distancing, KisslerScience2020}, i.e. $R_0 \rightarrow rR_0$, where 
\be
r:=(\bar{r}-1)s + 1.
\label{effect_trans}
\ee
Depending on the discrete or continuous nature of the intervention, we shall consider two kinds of such non-pharmaceutical policies. On one hand, we will speak of \emph{lockdown policies} when $s$ is only allowed to take the values $\{0, 1\}$; those correspond to situations in which a \textit{lockdown} is either on or off. For discrete $s(t)$ as in Eq.~\eqref{eq:s}, the effect of a lockdown ($s=1$) results in the reduction of the transmission rate $r = \bar{r}$ in Eq.~\eqref{effect_trans}. On the contrary, when the population is free to interact ($s=0$) then $r=1$ and there is no change in the disease's basic reproduction number.\\

\indent On the other hand, we will speak of \textit{physical-distancing policies} when $s$ is allowed to take any value in the interval $[0,1]$. Continuous values of $s(t)$ correspond to intermediate measures (for example mandatory face masks, suspension of sport events, remote working, school closures), the effect of which can be tentatively estimated from available data \cite{medical_parameters}. 

If distancing measures are the only type of intervention that a government uses, then a non-adaptive continuous policy is fully determined by the function $s(t; \bm{\mu})\in [0,1]$. To begin with, we will assume that the government can only declare new measures at the beginning of each week, i.e., that $s(t)$ does not vary within the weekly intervals $t\in [7k, 7(k+1)]=:I_k$, for $k\in\mathbb{Z}$. With these conditions, $s(t)$ can be expressed as 
\begin{equation}
s(t)=\sum_{k}\sigma(\tilde{s}_k)\chi_{I_k}(t).
\label{week_confi}
\end{equation}
where $\chi_{I_k}(t)$ is the characteristic function of week number $k$, i.e., $\chi_{I_k}(t)$ equals $1$ if $t$ is in the $k$-th week; and $0$, otherwise. The characteristic function, thus, ensures that there is only one policy $s(t)$ per week. $\sigma(y)=\frac{1}{1 + \mbox{exp}(y)}$ denotes the sigmoid function, which guarantees that $s(t)\in [0,1]$ by continuously mapping the variables $\{\tilde{s}_k\}_k$, such that, for $t\in[t_0,t_f]$, $s(t;\{s_k\})$ is everywhere differentiable, making it amenable to the gradient method. The parameters to optimize over are $\bm{\mu}\equiv\{\tilde{s}_k\}_k$, since they fully define the government's disease policy.

In 2021, several vaccines against COVID-19 successfully passed clinical trials, and nowadays many governments are fighting the disease through national vaccination campaigns. We model the effect of a COVID-19 vaccination campaign by assuming that: a) the government vaccinates the population across all compartments, i.e., without distinguishing between susceptible, exposed, recovered, etc.; b) just susceptible individuals benefit from the vaccine; c) all such vaccinated individuals become immune to the disease. The reader can find the explicit model in Appendix \ref{models_app}. Obviously, this model is a gross simplification of the effect of COVID-19 vaccines currently in the market, which on one hand varies depending on the particular vaccine and the prior exposure to COVID-19, and on the other hand does not seem to always confer full immunity to the disease \cite{vaccines}. More complicated vaccination models can be built to properly assess a government on vaccination policies, but, for the sake of illustration of our techniques, this simplified model will suffice.

We place a cap of $\Lambda=0.0011 \mbox{ days}^{-1}$ on the fraction of the population that can be vaccinated per day. We also assume that changes in the vaccination rate can only be made weekly. This brings us to parametrize the vaccination rate via the function

\begin{equation}
v(t)=\Lambda\sum_{k}\sigma(v_k)\chi_{I_k}(t).
\end{equation}
\noindent Finally, we assume that the government holds a supply of vaccines for just a third of the total population. This means that, together with (\ref{cc_constr}), we need to enforce the extra constraint

\be
\int_{t_0}^{t_f} dt v(t)\leq\frac{1}{3}.
\ee
\noindent We achieve this by adding the term (\ref{supply}) to the objective function, with $V=\frac{1}{3}$, $\lambda=10^4$. Without additional constraints, this model represent the case of a vaccine with 100\% efficacy. A reduced efficacy, let us denote it by $e<1$, can be straightforwardly modeled simply by multiplying the vaccination rate $v(t)$ and the vaccine supply $V$ by $e$.

In all our examples, we never allow the number of people in the critical care compartment to exceed the maximal occupancy, i.e., we impose the constraint in Eq.~\eqref{cc_constr}. To enforce it, we add the term (\ref{rho_C_c}) to the objective function, with $\rho(t)=\frac{100}{B_c}$.

Finally, in order to solve the differential equations (\ref{extra}), we use the Euler explicit method (\ref{Euler}) with step size $\delta=1.0$ to generate all our plots, apart from in Figures \ref{cc_ss_discrete_fig2} and \ref{vaccination_fig}, where we use $\delta=0.1$.

Next, we use the gradient method to derive the optimal interventions to control COVID-19 in a number of disease scenarios.
 
\subsection{Fighting COVID-19 via physical distancing measures}
We first consider policies exclusively based on non-pharmaceutical interventions; more concretely, continuous weekly policies $s(t)$ of the form (\ref{week_confi}). For starters, we tackle the problem of quickly steering the population towards herd immunity, all the while respecting the constraint (\ref{cc_constr}) on the critical care capacity. Our goal can be captured by a functional of the form (\ref{functional1cont}), with $\hat{A}(\bm{\mu})=0$ and

 \be 
 {\cal L}\left(t, \bm{\mu}, \bm{x}(t;\bm{\mu}, \bm{x_0})\right) = |S(t)-S_h|,
 \label{herd_func}
 \ee
 \noindent where $S$ (as depicted in Fig.~\ref{fig:schematic}) is the component of $\bm{x}$ that denotes the proportion of individuals in the population who are susceptible to the disease. $S_h=\frac{1}{R_0}$ is the proportion of susceptible individuals required for herd immunity to be guaranteed; thus, once $S(t) < S_h $, although people will continue to become infected, the natural evolution of the disease will be such that the rate and number of infected quickly dies out. 

\begin{figure}
   \centering
   \includegraphics[width=14 cm]{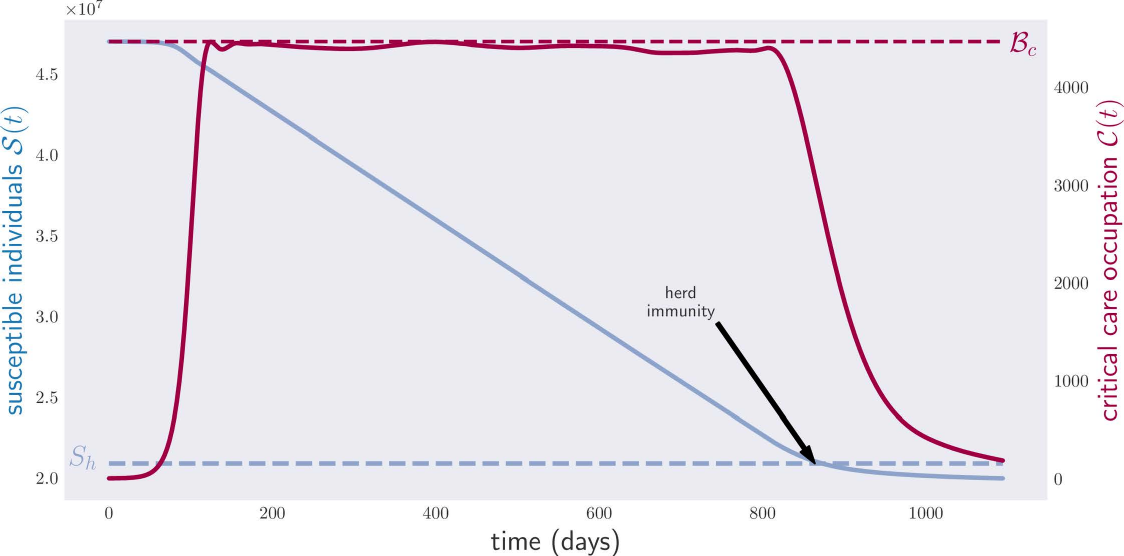}
   \caption{\textbf{Occupation of critical care beds (red) and population of susceptible individuals (blue) for a period of three years.} The blue line corresponds to the level of susceptibles guaranteeing herd immunity over the whole year.}
\label{cc_sus_herd_fig}
\end{figure}

 Fig.~\ref{cc_sus_herd_fig} illustrates the results of optimizing the objective function in Eq.~\eqref{herd_func} for continuous physical-distancing measures. The plot shows both the critical care occupancy ${\cal C}(t)$ and the total number ${\cal S}(t)$ of  susceptible individuals between times $t_0$ and $t_f=t_0+3\times 365$. The number of susceptible individuals reaches $S_h$ on day $864$, after which the disease can be considered extinct. As the reader can appreciate, the critical care occupancy curve never surpasses the critical value ${\cal B}_c$. 

\begin{figure}
  \centering
  \includegraphics[width=14 cm]{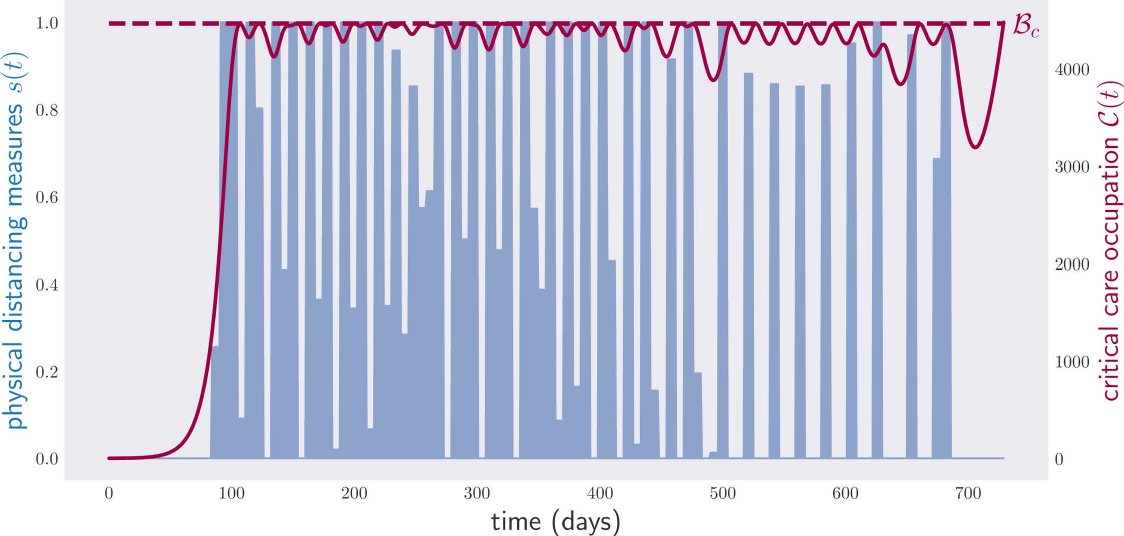}
  \caption{\textbf{Occupation of critical care beds (red) and physical-distancing measures (blue) for a period of two years.} The optimization has been performed over continuous weekly policies, $104$ continuous parameters, i.e., any value of physical-distancing measure $s$ between $0$ and $1$ is accepted. The algorithm, however, tends to prefer $0/1$ configurations, i.e., full lockdown or no lockdown in most instances. }
  \label{cc_ss_fig}
\end{figure}

The results in Figure \ref{cc_sus_herd_fig} are achieved at the cost of enforcing constant physical-distancing measures throughout the considered time frame. Naturally, the next problem we consider is that of minimizing the aggregate economic cost ${\cal E}$ associated with the physical-distancing measures implemented by the government over the first two years of the disease (while respecting the critical care capacity constraint (\ref{cc_constr})). Thus we take an objective function of the form \eqref{functional1cont}, with $\hat{A}(\bm{\mu})=0$ and
\be
{\cal L}\left(t, \bm{\mu}, \bm{x}(t;\bm{\mu}, \bm{x_0})\right) = {\cal E}(s(t;\bm{\mu}))
\,.
\label{economic_cost}
\ee
%
Figure \ref{cc_ss_fig} shows the result of applying the gradient method to minimize this functional for the cost function ${\cal E}(s)=s$.

This plot shows the critical care occupancy ${\cal C}(t)$, and also the physical-distancing measure $s(t)$ between times $t_0$ and $t_f=t_0+2\times 365$. The aggregate cost of the optimal policy is equal to the economic cost of sustaining a full lockdown for $294$ days. As anticipated, the optimal policy found by the computer is very complicated. Notice that the critical care occupancy grows quickly towards the end of the plot. The reason for this is that we asked the computer to minimize the total time in lockdown over a fixed period of two years, which is exactly what it did: it minimized the physical-distancing measures in the first two years, with complete disregard for what could happen next. 
To overcome such a pathological case, one can introduce additional terms into Eq.~\eqref{economic_cost} to make the final slope of the curve $\mathcal{C}(t)$ less steep, or even decreasing. 

\subsection{Fighting COVID-19 via lockdowns}
In some circumstances, the only distancing measures considered by governments are discrete: lockdown on or off, as in Eq.~\eqref{eq:s}. Let the objective function be given by eq. (\ref{economic_cost}), with ${\cal E}(s)=s$, i.e., we wish to minimize the total time under lockdown. As stated earlier, optimizations over discrete government interventions cannot be carried out directly with the gradient method. We conduct them instead with the two heuristics proposed in Appendix \ref{discrete_app}. 

\begin{figure}
  \centering
  \includegraphics[width=14 cm]{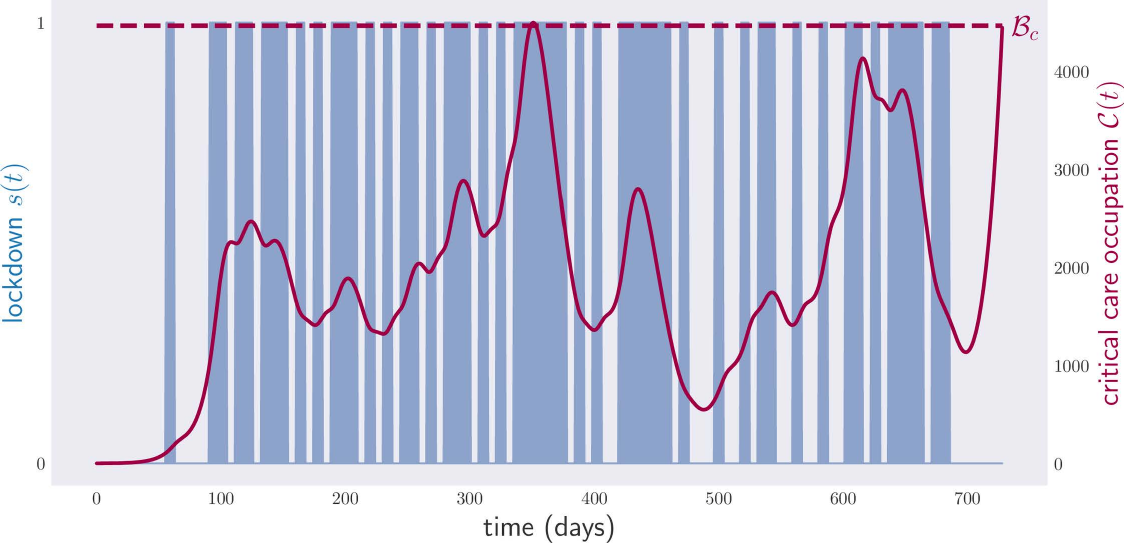}
  \caption{\textbf{Occupation of critical care beds (red) and lockdown (blue) for a period of two years.}  The plot shows the result of the optimization over probabilistic policies via gradient descent over a period of two years.}
  \label{cc_ss_discrete_fig}
\end{figure}

The first heuristic allows optimizing over weekly on/off confinements, and its results are shown in Figure \ref{cc_ss_discrete_fig}. This time the critical care occupancy curve touches the critical care capacity just once, after the end of year $1$. The reason for this is that we demanded lockdowns to last exactly one week: had we allowed the government to declare a lockdown on any day of the week, the computer would have found a tighter solution, with every peak of the red curve touching the dashed line. Even under this discrete weekly simplification, the solution found by the computer is non-trivial: it requires the government to declare a lockdown \emph{$27$ times} (as we will soon see, one can limit the total number of lockdowns in the final policy by adding constraints to the optimization problem). The total length of the lockdown in the course of two years is $371$ days.

\begin{figure}
  \centering
  \includegraphics[width=14 cm]{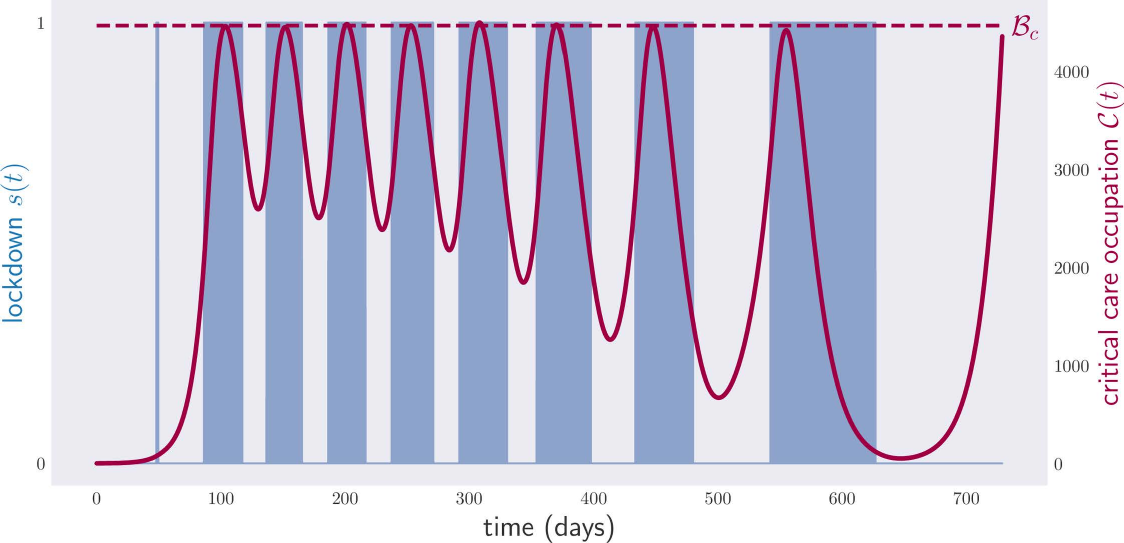}
  \caption{\textbf{Occupation of critical care beds (red) and lockdown measures (blue) for a period of two years.} Optimization over deterministic policies with arbitrary initial and final times for each lockdown period. A total of 9 lockdown periods has been fixed prior to the optimization. Notice that the first lockdown period, around day $60$, has been basically removed by the optimization procedure.}
  \label{cc_ss_discrete_fig2}
\end{figure}

In the second discretization method, the disease control policy is continuously parametrized through the vector $\bm{\mu}=(t_1, t_2,..., t_{2N})$, and lockdown is assumed to take place within the time intervals $[t_1, t_2], [t_3,t_4], ...$. In this parametrization, lockdowns can be declared or lifted at arbitrary times within $[t_0, t_f]$, and not only on Mondays, like in the first heuristic. This second discretization method has the advantage of allowing one to set the maximum number $N$ of lockdowns throughout the period $[t_0, t_f]$. For $N=9$, the corresponding critical care occupation and lockdown graphs are shown in Figure \ref{cc_ss_discrete_fig2}. The total length of the lockdown is $338$ days.

\subsubsection{Fighting COVID-19 via limited lockdowns and vaccination}
We next study the effect of vaccination campaigns in reducing the total time spent in lockdown. To ensure that the final recommended policies are easy to implement, we place the limit $N=5$ on the maximum number of lockdowns.

\begin{figure}
  \centering
  \includegraphics[width=14 cm]{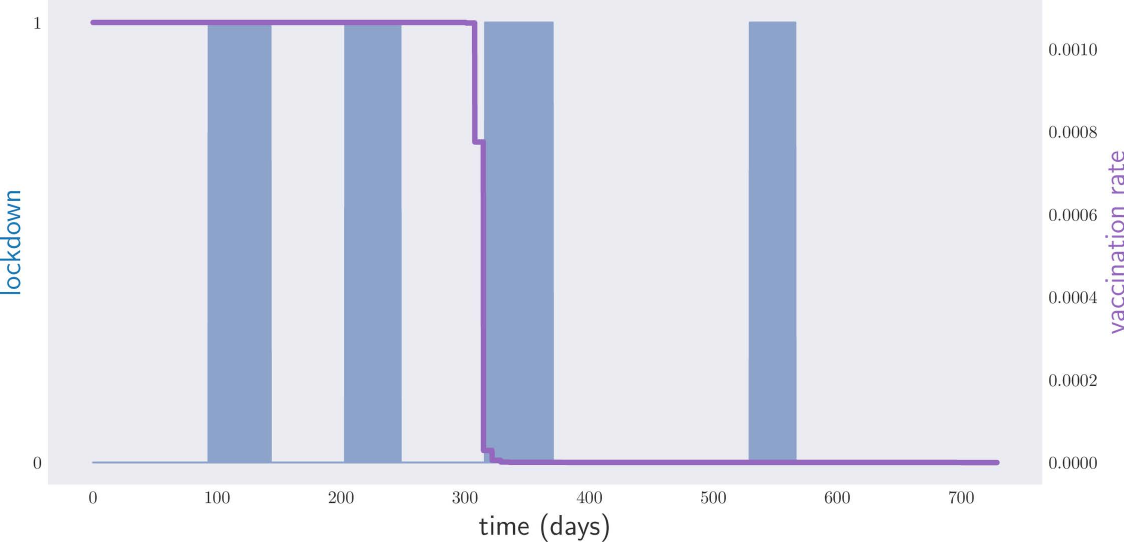}
  \caption{\textbf{Lockdown times (blue) and vaccination rates (magenta) for a period of two years.}}
  \label{vaccination_fig}
\end{figure}

If we allow the government to declare or lift lockdowns at any point in time, the second heuristic outputs the policy depicted in Figure \ref{vaccination_fig}. Note that the optimal policy does not require $5$ lockdowns, but $4$. The policy involves vaccinating the population at maximum rate since the very beginning of the government intervention until about day $200$, after which the vaccination rate gradually drops to zero. The total span of the required lockdown is $186$ days. This must be compared with the $338$-days lockdown required when we allow for $N=10$ lockdowns, but no vaccines are available. Even at such a slow vaccination pace, and under a shortage of supplies, the effects of a vaccination campaign prove to be very impressive.

\begin{figure}
  \centering
  \includegraphics[width=14 cm]{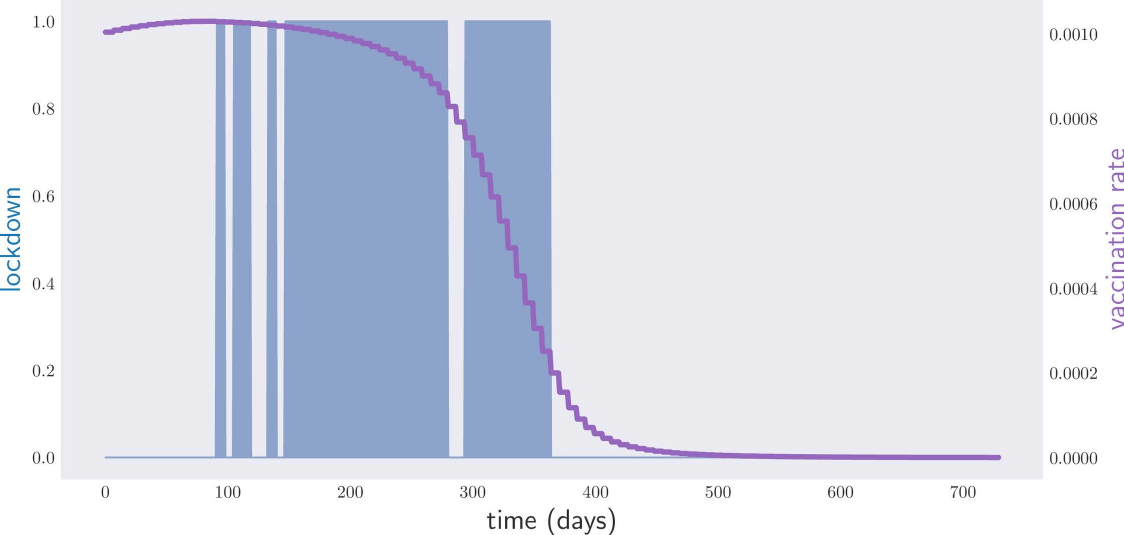}
  \caption{\textbf{Weekly lockdown times (blue) and vaccination rates (magenta) for a period of two years.}}
  \label{vaccination_fig2}
\end{figure}

Let us now see how the picture changes when the government is not allowed to enforce an intervention in the middle of the week. Enforcing a maximum number $N$ of lockdowns is not automatic when we deal with the first heuristic; it requires us to add an extra constraint to the optimization over non-deterministic policies. A possibility is to include the term

\be
\Omega\left\langle\left(\sum_j|c_j-c_{j+1}| - 2N\right)^+\right\rangle
\ee
\noindent in the objective function. Here the random variable $c_k$ equals $1$ if there was a lockdown on week $k$ or $0$ otherwise, see Appendix \ref{discrete_app}. The penalty for changing the government response by $2N+x$ times within the period $[t_0, t_f]$ is therefore $x\times\Omega$. Choosing $\Omega=10^2$, we arrive, through stochastic gradient descent, at the deterministic policy depicted in Figure \ref{vaccination_fig2}. It demands a total lockdown time of $231$ days (compared to the $371$ days of lockdown required by the corresponding non-pharmaceutical policy, depicted in Figure \ref{cc_ss_discrete_fig}).  As the reader can appreciate, the time spent in lockdown is considerably higher when we require lockdowns to be enforced or lifted on Mondays than when we allow the government to intervene at arbitrary times. 

\subsection{Dealing with uncertainty}

In practice, the predictions of any mathematical model for a physical system will not be perfect for a number of reasons. First, basic parameters of the model, such as the transmission rate or the initial occupation $\bm{x_0}$, are only known up to approximations. Even if reality were exactly described by a particular mathematical model, small errors in such parameters would accumulate in the long run, making long-term predictions unreliable. Second, reality is never exactly described by mathematical models: on the contrary, any tractable disease model is, at best, a rough approximation to reality. Consequently, even the most successful disease models in the market cease to deliver solid predictions beyond $4$ weeks \cite{predictability}.

\subsubsection{Uncertainty in the parameters}
These considerations make us question how practical a two-year disease control policy really is. Consider the physical-distancing policy depicted in Figure \ref{cc_ss_fig}, which was obtained by applying the gradient method to the SEIR model in \cite{social_distancing}. Here, the model parameters $\bm{\nu}$ correspond to the average values $\bm{\bar{\nu}}$ of the parameter ranges in Table \ref{params_tab} of Appendix \ref{models_app}. 
In reality of course, the values of the parameters are never all equal to their averages, so we proceed to generate sets of parameters with some fluctuations. Let $\Delta\bm{\nu}$ be the vector with entries given by the difference between the upper and lower bounds of all the entries of the table, and suppose that the actual parameters of ``reality'' are unknown and uniformly distributed in the region of values ${\cal N}_a=\{\nu:-a\frac{\Delta\bm{\nu}}{2}\leq \bm{\nu} -\bm{\bar{\nu}} \leq a\frac{\Delta\bm{\nu}}{2}\}$, where $a$ can be interpreted as the amount of noise or uncertainty. How robust is the afore-mentioned policy to uncertainty in the initial parameters?

\begin{figure}
  \centering
  \includegraphics[width=\textwidth]{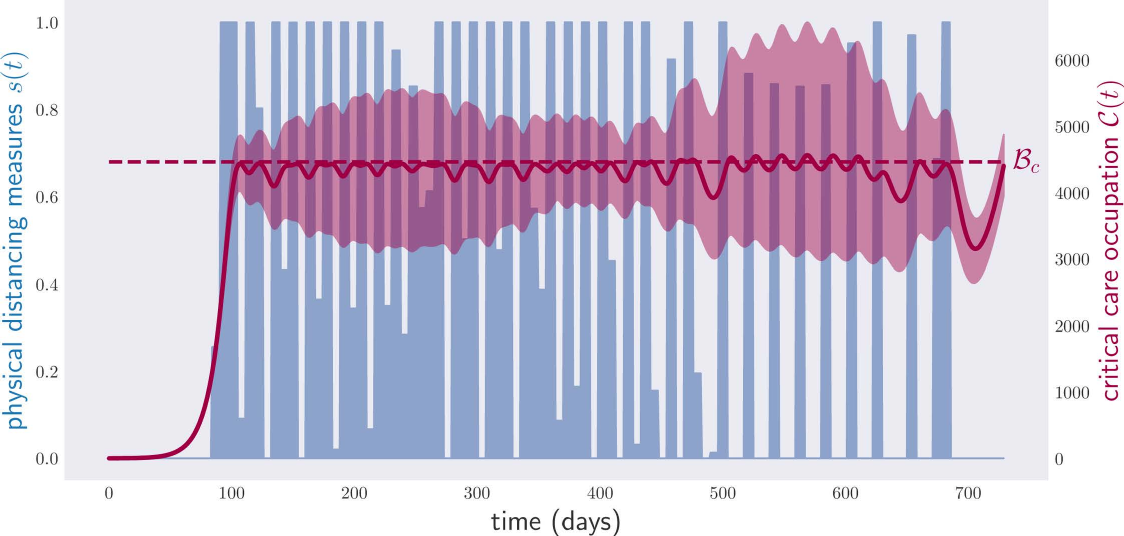}
  \caption{\textbf{Occupation of critical care beds (red) and (unoptimized) physical distancing measures (blue) for a period of two years under random parameters in ${\cal N}_{0.05}$.} The region in red is obtained by sampling $1000$ times from the region of model parameters ${\cal N}_{0.05}$ and evolving the corresponding models with the physical distancing policy optimized over the model with average-value parameters (as in Fig.~\ref{cc_ss_fig}). More precisely, the red region is the one delimited by the minimum and the maximum critical care occupation for all the $1000$ models, at each time. The red line represents the average critical care occupation in all those simulations. }
  \label{cc_ss_fig_noise}
\end{figure}

Fig.~\ref{cc_ss_fig_noise} shows the result of generating $1000$ independent parameter samples from the region ${\cal N}_{0.05}$, corresponding to a $5\%$ uncertainty, with respect to the given interval of values, and running the corresponding models for the optimal physical distancing policy in Fig.~\ref{cc_ss_fig}. 
As one can see, for some sampled values of parameters, the critical care capacity of the healthcare system is exceeded. This is not surprising, since the policy depicted in Fig.~\ref{cc_ss_fig} was devised to perform well under the assumption that $\bm{\nu}=\bm{\bar{\nu}}$ and not $\bm{\nu}\in {\cal N}_{0.05}$.

In order to tame the behavior in the plot in Fig.~\ref{cc_ss_fig_noise}, we use stochastic gradient descent to minimize the average value of the objective function $A$ assuming a uniform distribution of $\bm{\nu}$ over ${\cal N}_{a}$, as explained at the end of Appendix \ref{optimization_app}. By adding to $A$ sufficiently strong penalties for the violation of each optimization constraint, we make sure that such constraints will approximately hold for most of the points in ${\cal N}_{a}$.

\begin{figure}
  \centering
  \includegraphics[width=14 cm]{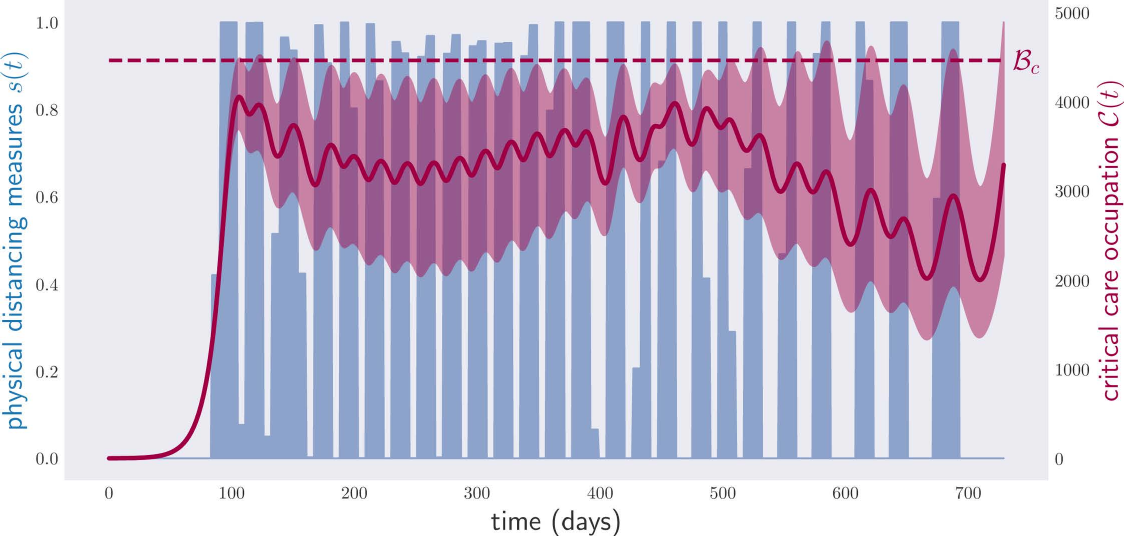}
  \caption{\textbf{Occupation of critical care beds (red) and (optimized) physical distancing measures (blue) for a period of two years under random parameters in ${\cal N}_{0.05}$.} The disease control policy was optimized to respect condition (\ref{cc_constr}) over the whole range of parameters ${\cal N}_{0.05}$. As in Fig.~\ref{cc_ss_fig_noise}, the region in red depicts again the range of critical care occupations observed in a sample of $1000$ model parameters in ${\cal N}_{0.05}$, and the red line the average critical care occupation.}
  \label{noise_optim}
\end{figure}

\begin{figure}
  \centering
  \includegraphics[width=14 cm]{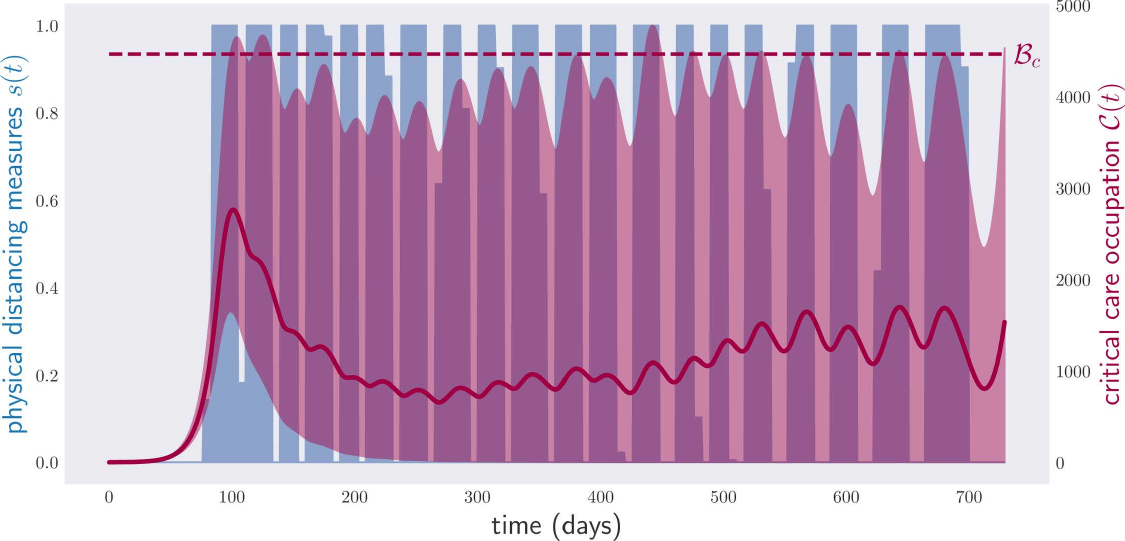}
  \caption{\textbf{Occupation of critical care beds (red) and (optimized) physical distancing measures (blue) for a period of two years under random parameters in ${\cal N}_{0.25}$.} The disease control policy was optimized to respect condition (\ref{cc_constr}) over the whole range of parameters ${\cal N}_{0.25}$. As in Fig.~\ref{cc_ss_fig_noise}, the region in red depicts again the range of critical care occupations observed in a sample of $1000$ model parameters in ${\cal N}_{0.25}$, and the red line the average critical care occupation.}
  \label{noise_optim25}
\end{figure}

Fig.~\ref{noise_optim} shows a lockdown policy minimizing the physical distancing measures under the condition that constraint (\ref{cc_constr}) holds for different values of $\nu\in {\cal N}_{0.05}$, i.e., the critical care capacity is not exceeded. This time, the violation of condition (\ref{cc_constr}) is neither so extreme nor so frequent. This comes, however, at the cost of enforcing physical distancing measures with a cost equivalent to $331$ days of lockdown. 
Repeating the optimization for ${\cal N}_{0.25}$, Fig.~\ref{noise_optim25}, we see that the critical care capacity is rarely surpassed. However, this time the total cost is equivalent to $414$ days of lockdown.

This result is what one would have expected. As time goes by, the predictions of the disease model for different values of $\bm{\nu}$ diverge: any policy that aims to satisfy constraint (\ref{cc_constr}) for large ranges of these parameters will necessarily require extensive physical distancing measures.

In practical policy-making, graphs such as Figs.~\ref{noise_optim} and \ref{noise_optim25} should not be understood to represent the actual physical distancing  policy, but rather to provide a provisional \emph{policy plan}. A policy plan gives a recommendation for action for the immediate future, given the current knowledge of the disease. In Fig.~\ref{noise_optim}, the policy plan is advising not to declare physical distancing measures in the first weeks. That is the measure that the government should adopt then. After a first time period, say four weeks, more data will have been gathered: this will allow us to obtain a better estimate of the parameters $\bm{\nu}$, and then re-run the models for another two years ahead. The measure to enforce should then be whatever the new policy plan recommends for the following four-week time period. The process is then repeated.

\begin{figure}
  \centering
  \includegraphics[width=14 cm]{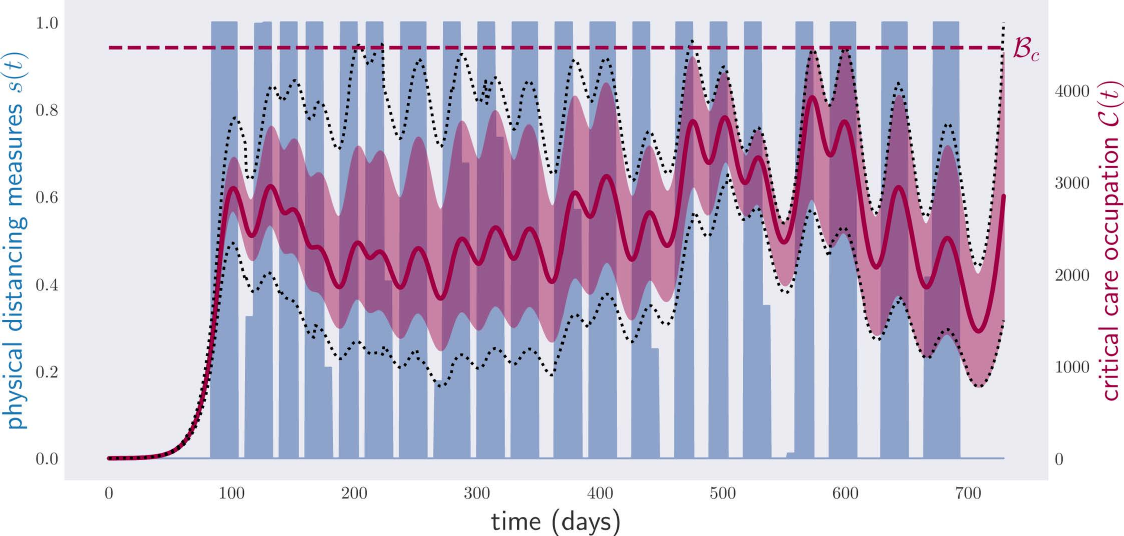}
  \caption{\textbf{Occupation of critical care beds (red) and physical distancing measures (blue) for a period of two years under random parameters and monthly noise decrease.} The disease control policy was optimized to respect the condition in Eq.~(\ref{cc_constr}), i.e., critical care capacity not exceeded, starting with the parameter region ${\cal N}_{0.25}$ and with a monthly noise decrease of $\frac{1}{\sqrt{k}}$, i.e., in the $k$-th month the noise is equal to $0.25/\sqrt{k}$. The black-dotted region depicts the range of critical care occupations observed in a sample of $1000$ model parameters in a sequence of shrinking regions ${\cal N}_{0.25/\sqrt{k}}$; more precisely, for each month $k$ the red region is obtained by evolving, from the initial time to month $k$ (included), $1000$ different models with parameters sampled from the region ${\cal N}_{0.25/\sqrt{k}}$. The final plot is obtained by joining the plots for each month $k$. The inner region in red depicts the range of critical care occupations corresponding to the uncertainty in the final month, i.e., obtained with $1000$ models with parameters sampled in ${\cal N}_{0.049}$. The red line represents the average critical care occupation, obtained by joining the average of the simulations with decreased uncertainty for each month $k$.  Despite the initial uncertainty on the parameters $\nu$ of the disease model, the final lockdown time is much lower, due to monthly revisions of the original policy plan.}
  \label{policy_plan}
\end{figure}

To test how this idea would perform in practice, we consider a scenario where the parameters defining the disease model are unknown, but the region in parameter space in which they live shrinks every month (to be precise, we used a $28$-day period, corresponding to four weeks). That is, at month $k$, the government is informed that the parameters $\bm{\nu}$ satisfy $\nu\in {\cal N}_{0.25/\sqrt{k}}$. Every four weeks, the policy is recalculated to minimize the physical distancing measures for the rest of the two years ahead, using the range $\bm{\nu}\in {\cal N}_{0.25/\sqrt{k}}$. The final curves for the critical care occupation and the physical distancing measures are shown in Fig. \ref{policy_plan} for $\bm{\nu}$, in a sequence of shrinking regions ${\cal N}_{0.25/\sqrt{k}}$, for each month $k$ (red region) and in the case of a fixed uncertainty region corresponding to the last month, i.e., ${\cal N}_{0.049}$ (inner dark blue region). The total cost of the physical distancing measures is equivalent to $358$ lockdown days. This has to be compared with the cost of $414$ days predicted by the initial policy plan under the assumption $\nu\in {\cal N}_{0.25}$.

In principle, one could further decrease the total planned cost by devising \emph{adaptive policy plans}, where the measure to be taken at each moment depends not only on the current time $t$, but also on the past history of physical distancing measures and their observed effects. In fact, we tried optimizing over generic adaptive policies described by a continuous version of a neural network architecture known as Long Short-Term Memory (LSTM) \cite{LSTM}. In all our numerical experiments, such simple LSTM architectures could not improve the performance of non-adaptive strategies, but this could be due to ineffective training on our side.

\subsubsection{Nondeterministic models}

\begin{figure}
  \centering
  \includegraphics[width=14 cm]{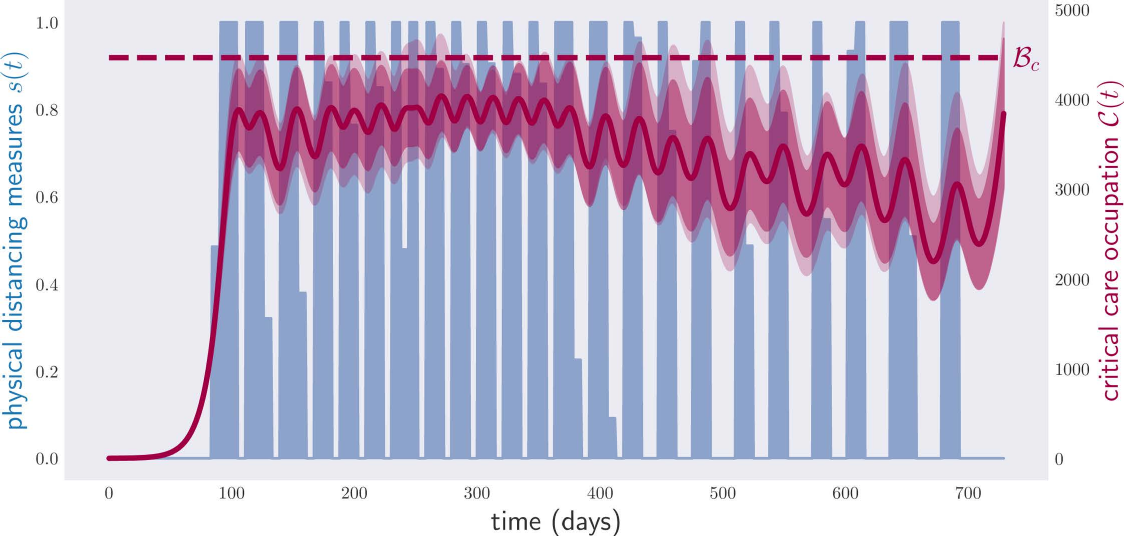}
  \caption{\textbf{Occupation of critical care beds (red) and physical distancing measures (blue) for a period of two years for stochastic evolution with noise-level $\eta=0.05$.} The disease control policy was optimized to respect the condition in Eq.~(\ref{cc_constr}), i.e., critical care capacity not exceeded. The region in light red depicts the range of critical care occupations observed in a sample of $1000 \times \eta$ different evolutions, with a stochastic noise $\eta=0.05$; the red line, the average critical care occupation; and the darker red region, the trajectories that lie within two standard deviations from the average value.  The policy corresponds to a total of 321 lockdown days.}
  \label{stoch_evo5}
\end{figure}

So far, all models considered in the optimization are deterministic, namely, given the model parameters and initial conditions, the compartments evolve along a unique trajectory. As a more general case, we can consider the one in which the disease's dynamics is governed, rather than by Eq.\eqref{harvard_model}, by a stochastic differential equation. More concretely, consider the model that results when we replace the first and second lines of  Eq.\eqref{harvard_model} by
\begin{align}
&\frac{dS}{dt} = -(1+\xi)r(t)\beta(t) S (I_R+I_H+I_C), \nonumber\\
&\frac{dE}{dt} = (1+\xi)r(t)\beta(t) S (I_R+I_H+I_C) -\nu E,
\label{eq:stoch_diffeq}
\end{align}
\noindent where $\xi$ represents a Gaussian noise term reflecting stochastic fluctuations on the virus' transmissivity~\cite{book}. Solving this equation by discretization, each occurrence of $\xi$ at time $t_k = t_0 + k \delta$ is replaced by a quantity $\xi_k$ sampled from a Gaussian distribution of zero mean and standard deviation $\eta\sqrt{\delta}$, where $\eta$ denotes the magnitude of the noise term. 

For our simulation, we consider the noise $\eta=0.05$. As before, the method of stochastic gradient descent is used for obtaining a noise-robust policy. At each time step, we average the gradient over  $1024 \times \eta$  different stochastic evolutions. In contrast to the case of noisy parameters, for a stochastic evolution the sampling has to be repeated at each time step, see Appendix~\ref{optimization_app} for more details. The results of our simulations are shown in Fig.~\ref{stoch_evo5}. Notice that, in contrast to Fig.~\ref{cc_ss_fig_noise}, the noise is not uniformly distributed in a limited interval, but is given by a Gaussian, which allows for (rare) events far away from the mean. This justifies the different representations of uncertainty in these plots. Despite the presence of relatively high noise in the evolution, as shown by the wide fluctuations in Fig.~\ref{stoch_evo5}, the algorithm is able to devise strong policies that maintain the critical care occupation almost always, i.e., with a high probability, below its capacity.

\section{Conclusion}
In this paper, we have applied standard tools from optimization theory and machine learning to identify optimal disease control policies, given an epidemiological model. This is in stark contrast to standard practice in mathematical epidemiology, where human intuition is used to narrow down the considered set of policies to a uni-parametric family. We saw that the optimal solutions found by our algorithms are highly counter-intuitive, and thus unlikely to be identified by a human. This supports the idea that policies for disease control should be based on a combination of both human expertise and machine learning.

Compared to previous approaches that tried to identify suitable disease control policies through optimal control theory, our framework allows one to devise policies that satisfy arbitrary constraints under arbitrary uncertainties in the initial conditions and model parameters that determine the disease's dynamics, and stochastic evolution. Our methods, in addition, allow one to optimize over discrete policies, as well as policies that can just vary at certain fixed times.

To illustrate our ideas, we studied a scenario in which a computer is tasked with outputting the minimal amount of non-pharmaceutical interventions for an epidemic in a hypothetical country, in such a way as to never exceed the critical care bed capacity. We looked at situations in which these measures were continuous (recommendations on the interval $[0, 1]$) as well as discrete (either $0$ or $1$) -- a lockdown that is off or on, respectively for periods of $2$ years. We experimented with measures that are just allowed to change weekly, as well as those in which there is a maximum number of lockdowns that is allowed to be declared. We also showed how vaccine supplies, even when meager and poorly distributed, can dramatically shorten the total confinement required to keep the disease under control.

Admittedly, some of our computer-generated policies are too complicated to be implemented in practice. Our formalism allows, however, to limit the complexity of the found solutions by adding extra constraints to the original optimization problem. To highlight this feature, we generated a near-optimal weekly policy plan of vaccinations and on/off weekly confinement measures with a cap on the total number of lockdowns.

We examined the problems that one may encounter when applying our techniques to scenarios where the model parameters are not known with high accuracy, or the model evolves stochastically. This led us to propose practical policy plans which must be continually revisited, to account for our ever-changing and ever-growing knowledge in an epidemic. We tested the viability of this approach by simulating a scenario where the uncertainty on the disease model parameters decreases with time. As expected, the final policy implemented was safe for the final range of parameters and required considerably less physical distancing measures than the initial policy plan hinted. 

In this last regard, an interesting problem for future research is devising a gradient-friendly ansatz for adaptive policy plans for disease control, where the actual measure at each time depends on the whole history of disease indicators accessible to the government. In theory, such plans should predict lower values of the average objective function in scenarios where the model parameters are unknown. In our experience, though, gradient descent applied to the standard LSTM architecture seems to be unable to beat the non-adaptive score.

Finally, we would like to remark once more that, since the optimization problems we dealt with in this paper are non-convex, the gradient method is not guaranteed to converge to the minimum of the (average) objective function. While conducting this research, in order to convince ourselves that the solutions found by our numerical methods were close to optimal, we had to repeat our optimizations several times, with different initial policies $\bm{\mu^{(0)}}$ and learning rates $\epsilon$. Such a redundant use of computational resources would have been entirely avoidable if we had had some rough approximation to the exact solution of the problem. Hence we conclude this paper with a challenge for the operations research community: develop mathematical tools which allow one to \emph{lower bound} the solution of minimization problems involving ordinary differential equations.

\section*{Acknowledgments} 
We thank Luca Gerardo-Giorda and Mario Budroni for useful discussions.

\bibliography{corona}

\begin{thebibliography}{10}

\bibitem{un_framework}
Nations U. A UN framework for the immediate socio-economic response to
  COVID-19; 2020.
\newblock Available from:
  \url{https://unsdg.un.org/resources/un-framework-immediate-socio-economic-response-covid-19}.

\bibitem{yemen}
COVID-19 has made the health system's collapse complete in {Y}emen.
\newblock Medecins sans Frontieres. 2020;.

\bibitem{salvador}
People are dying at home amid collapsing health system in {E}l {S}alvador.
\newblock Medecins sans Frontieres. 2020;.

\bibitem{kathmandu}
Pradhan TR.
\newblock Government decides to lift the four-month-long coronavirus lockdown,
  but with conditions.
\newblock The Kathmandu Post. 2020;.

\bibitem{book}
Keeling MJ, Rohani P.
\newblock Modeling Infectious Diseases in Humans and Animals.
\newblock Princeton University Press; 2008.
\newblock Available from: \url{http://www.jstor.org/stable/j.ctvcm4gk0}.

\bibitem{analytic_discrete}
Hindes J, Bianco S, Schwartz I.
\newblock Optimal periodic closure for minimizing risk in emerging disease
  outbreaks.
\newblock PLoS ONE. 2021;16(1 January).
\newblock doi:{10.1371/journal.pone.0244706}.

\bibitem{control_theory1}
Gaff H, Schaefer E.
\newblock Optimal control applied to vaccination and treatment strategies for
  various epidemiological models.
\newblock Mathematical Biosciences and Engineering. 2009;6(1551-0018 2009 3
  469):469.
\newblock doi:{10.3934/mbe.2009.6.469}.

\bibitem{control_theory2}
de~Pinho MdR, Kornienko I, Maurer H.
\newblock Optimal Control of a SEIR Model with Mixed Constraints and L1 Cost.
\newblock In: Moreira AP, Matos A, Veiga G, editors. CONTROLO'2014 --
  Proceedings of the 11th Portuguese Conference on Automatic Control. Cham:
  Springer International Publishing; 2015. p. 135--145.

\bibitem{control_theory3}
Obsu LL, Balcha SF.
\newblock Optimal control strategies for the transmission risk of COVID-19.
\newblock Journal of Biological Dynamics. 2020;14(1):590--607.
\newblock doi:{10.1080/17513758.2020.1788182}.

\bibitem{control_theory4}
Zamir M, Shah Z, Nadeem F, Memood A, Alrabaiah H, Kumam P.
\newblock Non Pharmaceutical Interventions for Optimal Control of COVID-19.
\newblock Computer Methods and Programs in Biomedicine. 2020;196:105642.
\newblock doi:{https://doi.org/10.1016/j.cmpb.2020.105642}.

\bibitem{R_0}
Li Q, Guan X, Wu P, Wang X, Zhou L, Tong Y, et~al.
\newblock Early Transmission Dynamics in Wuhan, China, of Novel
  Coronavirus-Infected Pneumonia.
\newblock New England Journal of Medicine. 2020;382(13):1199--1207.
\newblock doi:{10.1056/NEJMoa2001316}.

\bibitem{R_0_bis}
Riou J, Althaus CL.
\newblock Pattern of early human-to-human transmission of Wuhan 2019 novel
  coronavirus (2019-nCoV), December 2019 to January 2020.
\newblock Eurosurveillance. 2020;25:2000058.
\newblock doi:{https://doi.org/10.2807/1560-7917.ES.2020.25.4.2000058}.

\bibitem{lockdown}
Volpicelli G.
\newblock China has almost eliminated Covid-19. What can the world learn?
\newblock Wired. 2020;.

\bibitem{deep_learning}
Goodfellow IJ, Bengio Y, Courville A.
\newblock Deep Learning.
\newblock Cambridge, MA, USA: MIT Press; 2016.
\newblock Available from: \url{http://www.deeplearningbook.org}.

\bibitem{social_distancing}
Kissler S, Tedijanto C, Lipsitch M, Grad Y. Social distancing strategies for
  curbing the {COVID}-19 epidemic; 2020.
\newblock Available from: \url{https://dash.harvard.edu/handle/1/42638988}.

\bibitem{predictability}
Reich NG, Brooks LC, Fox SJ, Kandula S, McGowan CJ, Moore E, et~al.
\newblock A collaborative multiyear, multimodel assessment of seasonal
  influenza forecasting in the {U}nited {S}tates.
\newblock Proceedings of the National Academy of Sciences.
  2019;116(8):3146--3154.
\newblock doi:{10.1073/pnas.1812594116}.

\bibitem{githuburl}
Online repository with the code used for the simulations, optimizations, and
  plots presented;.
\newblock Available from:
  \url{https://github.com/costantinobudroni/opt-disease-control/}.

\bibitem{open_systems}
Keeling MJ, Rohani P.
\newblock Estimating spatial coupling in epidemiological systems: a mechanistic
  approach.
\newblock Ecology Letters. 2002;5(1):20--29.
\newblock doi:{10.1046/j.1461-0248.2002.00268.x}.

\bibitem{WHO_Press2020}
World {H}ealth {O}rganization, {P}ress briefing, {M}arch 20th 2020;.
\newblock Available from:
  \url{https://www.who.int/docs/default-source/coronaviruse/transcripts/who-audio-emergencies-coronavirus-press-conference-full-20mar2020.pdf?sfvrsn=1eafbff_0}.

\bibitem{vaccination_SEIR}
McLean AR, Blower SM.
\newblock Imperfect vaccines and herd immunity to {HIV}.
\newblock Proceedings of the Royal Society of London Series B: Biological
  Sciences. 1993;253(1336):9--13.
\newblock doi:{10.1098/rspb.1993.0075}.

\bibitem{pulse_vaccination}
Agur Z, Cojocaru L, Mazor G, Anderson RM, Danon YL.
\newblock Pulse mass measles vaccination across age cohorts.
\newblock Proceedings of the National Academy of Sciences.
  1993;90(24):11698--11702.
\newblock doi:{10.1073/pnas.90.24.11698}.

\bibitem{foot_and_mouth}
Keeling MJ, Woolhouse MEJ, Shaw DJ, Matthews L, Chase-Topping M, Haydon DT,
  et~al.
\newblock Dynamics of the 2001 {UK} Foot and Mouth Epidemic: Stochastic
  Dispersal in a Heterogeneous Landscape.
\newblock Science. 2001;294(5543):813--817.
\newblock doi:{10.1126/science.1065973}.

\bibitem{subgradient}
Boyd S, Xiao L, Mutapcic A.
\newblock Subgradient methods.
\newblock lecture notes of EE392o, Stanford University, Autumn Quarter. 2004;.

\bibitem{stoch}
Duchi J.
\newblock EE364b: Lecture Slides and Notes.
\newblock \texttt{https://webstanfordedu/class/ee364b/lectureshtml}. 2018;.

\bibitem{support}
Cristianini N, Shawe-Taylor J.
\newblock An Introduction to Support Vector Machines and Other Kernel-based
  Learning Methods.
\newblock Cambridge University Press; 2000.

\bibitem{raccoons}
Keller JP, Gerardo-Giorda L, Veneziani A.
\newblock Numerical simulation of a susceptible-exposed-infectious
  space-continuous model for the spread of rabies in raccoons across a
  realistic landscape.
\newblock Journal of biological dynamics. 2013;7 Suppl 1:31—46.
\newblock doi:{10.1080/17513758.2012.742578}.

\bibitem{adam}
Kingma DP, Ba J.
\newblock Adam: {A} Method for Stochastic Optimization.
\newblock In: Bengio Y, LeCun Y, editors. 3rd International Conference on
  Learning Representations, {ICLR} 2015, San Diego, CA, USA, May 7-9, 2015,
  Conference Track Proceedings; 2015.Available from:
  \url{http://arxiv.org/abs/1412.6980}.

\bibitem{SAT}
Cook SA.
\newblock The Complexity of Theorem-Proving Procedures.
\newblock In: Proceedings of the Third Annual ACM Symposium on Theory of
  Computing. STOC ’71. New York, NY, USA: Association for Computing
  Machinery; 1971. p. 151–158.
\newblock Available from: \url{https://doi.org/10.1145/800157.805047}.

\bibitem{medical_parameters}
et~al F.
\newblock Report 9: Impact of non-pharmaceutical interventions ({NPI}s) to
  reduce {COVID}-19 mortality and healthcare demand.
\newblock Imperial College. 2020;.

\bibitem{KisslerScience2020}
Kissler SM, Tedijanto C, Goldstein E, Grad YH, Lipsitch M.
\newblock Projecting the transmission dynamics of {SARS-CoV}-2 through the
  postpandemic period.
\newblock Science. 2020;368(6493):860--868.

\bibitem{vaccines}
Katella K. Comparing the COVID-19 Vaccines: How Are They Different?; 2021.
\newblock Available from:
  \url{https://www.yalemedicine.org/news/covid-19-vaccine-comparison}.

\bibitem{LSTM}
Hochreiter S, Schmidhuber J.
\newblock Long Short-term Memory.
\newblock Neural computation. 1997;9:1735--80.
\newblock doi:{10.1162/neco.1997.9.8.1735}.

\bibitem{numerics}
Front Matter.
\newblock John Wiley \& Sons, Ltd; 2016.
\newblock Available from:
  \url{https://onlinelibrary.wiley.com/doi/abs/10.1002/9781119121534.fmatter}.

\bibitem{fem}
Logan DL.
\newblock A First Course in the Finite Element Method Using Algor.
\newblock 2nd ed. USA: Brooks/Cole Publishing Co.; 2000.

\end{thebibliography}

\begin{appendix}

\section{Models for the spread of COVID-19}
\label{models_app}

\begin{figure}
  \centering
  \includegraphics[width=14 cm]{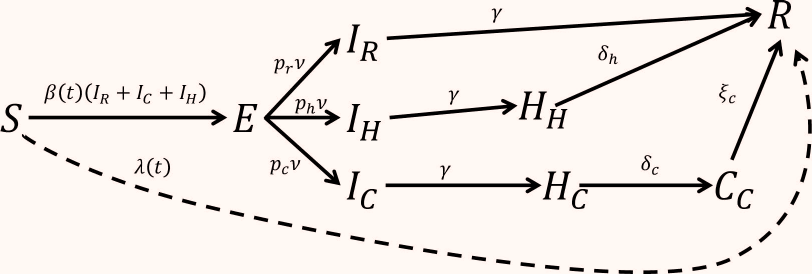}
  \caption{\textbf{Two COVID-19 compartmental models for disease policies based on physical distancing measures.} The solid lines show the model first presented in \cite{KisslerScience2020}, where the coefficients connecting the compartments are interpreted couplings, population fractions, time delays and so forth. The addition of the dotted line connecting the `susceptible' and `recovered' compartments creates another model where a vaccination campaign is taken into account. }
  \label{harvard_model_fig}
\end{figure}

In all our numerical simulations, we assume that the dynamics of the COVID-19 are well approximated by a compartmental model of the SEIR type. When the government policy reduces to enforcing physical distance measures, we adopt a simplified version of the model used in \cite{social_distancing}. This model divides those infected with COVID-19 into three different compartments: $I_R$, or those who recover by themselves from the disease; $I_H$, those who require hospitalization but do not enter a critical care unit; and $I_C$, those who are both hospitalized and visit a critical care unit before recovery. The dynamics of the model are governed by the system of ordinary differential equations below, see the diagram in Figure \ref{harvard_model_fig}:
\begin{align}
&\frac{dS}{dt} = -r(t)\beta(t) S (I_R+I_H+I_C), \nonumber\\
&\frac{dE}{dt} = r(t)\beta(t) S (I_R+I_H+I_C) -\nu E,\nonumber\\
&\frac{dI_R}{dt} = p_R\nu E-\gamma I_R, \nonumber\\
&\frac{dI_H}{dt} = p_H\nu E -\gamma I_H, \nonumber\\
&\frac{dI_C}{dt} = p_C\nu E -\gamma I_C, \nonumber\\
&\frac{dH_H}{dt}=\gamma I_H-\delta_H H_H, \nonumber\\
&\frac{dH_C}{dt}=\gamma I_C-\delta_C H_C, \nonumber\\
&\frac{dC}{dt}=\delta_C H_C-\xi C, \nonumber\\
&\frac{dR}{dt}=\gamma I_C + \delta_H H_H + \xi C.
\label{harvard_model}
\end{align}
Here 
\be
\beta(t)=\gamma R_0\left(\frac{1+\Delta}{2}+\frac{1-\Delta}{2}\cos\left(\frac{2\pi (t+\phi)}{7\times 52}\right)\right)
\ee
\noindent denotes the virus' transmitivity, that is subject to seasonal variability. $r(t)=(\bar{r}-1)s(t)+1$ models the effect of a government-mandated lockdown $s(t)\in\{0,1\}$ on the virus' transmission rate. The values of the remaining parameters are taken from \cite{social_distancing}, and appear in Table \ref{params_tab}.
\begin{table}
\centering
\begin{tabular}{|c|c|c|}
\hline
parameter&value&units\\ \hline
$\bar{r}$&$[0, 0.6]$&none\\
$\Delta$&$[0.7, 1.0]$&none\\
$\gamma$&$1/5$& days$^{-1}$\\
$\nu$&$1/4.6$& days$^{-1}$\\
$p_R$&$0.9596$& none\\
$p_H$&$0.0308$&none\\
$p_C$&$0.0132$&none\\
$\delta_H$&$1/8$& days$^{-1}$\\
$\delta_C$&$1/6$& days$^{-1}$\\
$\xi$&$1/10$& days$^{-1}$\\
$\phi$ & $-7\times 3.8$& days\\
$R_0$ &[2, 2.5]&None\\
\hline
\end{tabular}
\caption{Parameter ranges for the compartmental model for COVID-19 proposed in \cite{social_distancing}.}
\label{params_tab}
\end{table}

If the government has a vaccine available, the model changes. In that case, the first and last lines of eq. (\ref{harvard_model}) shall be replaced by

\begin{align}
&\frac{dS}{dt} = -r(t)\beta(t) S (I_R+I_H+I_C)-v(t)S, \nonumber\\
&\frac{dR}{dt}=\gamma I_C + \delta_H H_H + \xi C+v(t)S,
\label{harvard_model_vacc}
\end{align}
\noindent where $v(t)$ denotes the vaccination rate.

In all our numerical simulations, we take the total population to be $47$ million; the critical care bed capacity per inhabitant $C_c$ is also taken to be $9.5\times 10^{-5}$. We assume that the government starts its intervention on day $t_0=60$, $30$ days after the outbreak of the disease. We model the disease outbreak by assuming that, at time $t_{out}=30$, there are $10$ individuals in compartment $E$. By default, the values of the disease parameters are taken to be the arithmetic means of the intervals shown in Table \ref{params_tab}. We assume that the government can vaccinate at most $50,000$ individuals per day. This means that, at any given time $t$, $v(t)\leq\Lambda:=\frac{50,000}{47,000,000}\approx 0.00106$. In addition, we assume that the total supply of vaccines can just cover a third of the total population. In other words, $\int dt v(t)\leq \frac{1}{3}$.

\section{Optimization over continuous measures for disease control}
\label{optimization_app}

In this section, we explain how to apply the gradient method to optimize over continuous classes of government interventions. Our starting point is an ordinary differential equation of the form
\be
\frac{dx^i}{dt} = G^i(t,\bm{x};\bm{\mu});
\quad i=1,...,m.
\label{original}
\ee
The entries of vector $x$ represent the occupations of the different compartments of a disease model. In the case of adaptive policies, some of such entries might also represent the components of the \emph{cell state} $\bm{\theta}$ \cite{LSTM}, i.e., the internal variables used by the government to keep track of the evolution of the disease and guide future government interventions. $\bm{\mu}\in \R^n$ represents a parametrization of the effects of a given policy. Let $\bm{x}(t;\bar{\bm{\mu}},\bm{x}_0)$ be the solution of Eq. (\ref{original}) with initial conditions $\bm{x}(0)=\bm{x}_0$ and $\bm{\mu}=\bar{\bm{\mu}}$.

Given the set $M\subset \R^n$, we consider the problem of finding the parameters $\bm{\mu}^\star\in M$ such that $\bm{x}(t;\bm{\mu}^\star,\bm{x}_0)$ minimizes a given functional $A$. This functional defines the means by which we wish to control the disease: it might represent the number and duration of lockdown, etc. For the time being, let us assume this functional to be of the form 

\be
A(\bm{\mu},\bm{x}_0) =\int_{t_0}^{t_f} dt
\;
{\cal L}(t, \bm{\mu}, \bm{x}(t;\bm{\mu},\bm{x}_0)) 
+ \hat{A}(\bm{\mu})
\,,
\label{functional1}
\ee
\noindent From the discussion in Section~\ref{overview}, the functional in Eq.~\eqref{functional1} might also contain constraints which we wish the solution $\bm{x}(t;\bm{\mu}^\star,\bm{x}_0)$ to satisfy, such as (\ref{rho_C_c}), (\ref{supply}). Note that we are assuming to know the initial conditions $\bm{x}_0$ with precision. We will relax this requirement by the end of the section.

To minimize $A(\bm{\mu},\bm{x}_0)$ via gradient descent, we need to compute $\nabla_{\bm{\mu}} A$. For functionals of the form (\ref{functional1}), we have that
\be
\frac{\partial A}{\partial \mu_j}=\int_{t_0}^{t_f} dt \left(\frac{\partial {\cal L}(t, \bm{\mu}, \bm{x}(t;\bm{\mu}, \bm{x}_0))}{\partial \mu_j}+ \sum_i \frac{\partial {\cal L}(t, \bm{\mu}, \bm{x}(t;\bm{\mu}, \bm{x}_0))}{\partial x^i}\frac{\partial x^i}{\partial \mu_j}\right)+\frac{\partial \hat{A}(\bm{\mu})}{\partial\mu_j}.
\label{partial}
\ee

The next question is thus how to compute the derivatives $\frac{\partial x^i}{\partial \mu_j}$. To this aim, define the variables $y^i_j(t; \bm{\mu},\bm{x}_0)\equiv \frac{\partial x^i(t; \bm{\mu},\bm{x}_0)}{\partial\mu_j}$. Differentiating equation (\ref{original}) by $\mu_j$, we have that

\be
\frac{\partial y^i_j}{\partial t}=\frac{\partial G^i(t,\bm{x};\bm{\mu})}{\partial\mu_j} + \sum_l \frac{\partial G^i(t, \bm{x};\bm{\mu})}{\partial x^l}y^l_j; i=1,...,m, j =1,...,n.
\label{extra}
\ee

In order to obtain $\{y^i_j(t)\}$ for each time $t$, it hence suffices to solve the system of coupled differential equations given by (\ref{original}), (\ref{extra}) with initial conditions $\bm{x}(t_0)=\bm{x}_0$, $y^i_j(t_0)=0$. This can be achieved numerically through several different methods, depending on the desired accuracy. The simplest such method is called \emph{Euler explicit} \cite{numerics}: for some $\delta>0$, it consists of regarding time as a discrete variable of the form $t_k=t_0 + \delta k$, for $k=0,..., \lceil \frac{t_f-t_0}{\delta}\rceil$. The quantities $\{x^i(t_k), y^i_j(t_k):k\}$ are then obtained by recursively applying the relations

\begin{align}
&x^i(t_{k+1}) = x^i(t_{k}) + \delta G^i(t_k, \bm{x}(t_k); \bm{\mu}),\nonumber\\
&y_j^i(t_{k+1}) = y_j^i(t_k) + \delta\left(\frac{\partial G^i(t_k, \bm{x}(t_k); \bm{\mu})}{\partial \mu_j} + \sum_l\frac{\partial G^i(t_k, \bm{x}(t_k);\bm{\mu})}{\partial x^l}y^l_j(t_k)\right).
\label{Euler}
\end{align}

We will also encounter situations where our functional $A$ is more complicated than (\ref{functional1}). Some parameters $\bm{\zeta}$ (not policy parameters) regulating the evolution (\ref{original}), such as the basic reproduction number of the disease, might be unknown, or perhaps the initial conditions $\bm{x}_0$ are just known within some bounds. In such cases, the problem's objective function $A$ might adopt the form
\be
A(\bm{\mu}) = \int p(\bm{\zeta},\bm{x}_0)d\bm{\zeta} d\bm{x}_0\int dt {\cal L}(t,\bm{\mu}, \bm{x}(t;\bm{\mu},\bm{\zeta}, \bm{x}_0)) +\hat{A}(\mu),
\label{functional2}
\ee
\noindent for some probability measure $p(\bm{\zeta},\bm{x}_0)d\bm{\zeta} d\bm{x}_0$. Again, we wish to minimize $A$ over $\bm{\mu}$. As explained in section \ref{gradient_sec}, this can be achieved via stochastic gradient descent methods \cite{subgradient}: all we need is an unbiased estimator $\tilde{\nabla}_{\bm{\mu}}A$ of $\nabla_{\bm{\mu}}A$. We obtain this estimator by taking $N$ independent samples $(\bm{\zeta}^{(j)}, \bm{x}_0^{(j)})_{j=1}^N$ from the measure $p(\bm{\zeta},\bm{x}_0)d\bm{\zeta} d\bm{x}_0$ and using them to compute the quantity
\be
\tilde{\nabla}_{\bm{\mu}}A =\nabla_{\bm{\mu}}\hat{A}(\mu)+\frac{1}{N}\sum_{j=1}^N \nabla_{\bm{\mu}}\int dt {\cal L}(t, \bm{\mu}, \bm{x}(t;\bm{\mu}, \bm{\zeta}^{(j)}, \bm{x}_0^{(j)})).
\label{estimator_grad}
\ee

The prescription above also allows optimizing over control policies in scenarios where the disease's equations of motion are not deterministic, but probabilistic. Suppose, for instance, that the disease's dynamics are governed by a stochastic differential equation

\be
\frac{dx^i}{dt} = G^i(t,\bm{x};\bm{\mu},\bm{\xi}); i=1,...,m,
\label{stoch_diff}
\ee
\noindent where the entries of $\bm{\xi}\in \R^s$ represent Gaussian white noise. When we solve this equation by discretization, each occurrence of $\bm{\xi}$ at time $t_k=t_0+k\delta$ is to be replaced by an $s$-dimensional vector $\bm{\xi}^{k}$ of independent Gaussian variables of $0$ mean and standard deviation $\sqrt{\delta}$. Call $M=\lceil \frac{t_f-t_0}{\delta}\rceil$. If we aim to minimize the expectation value of the objective function, we can estimate the gradient of the average through eq. (\ref{estimator_grad}), with $\bm{\zeta}=(\bm{\xi}^{1}, ..., \bm{\xi}^{M})$ and

\be
p(\bm{\zeta})=\prod_{k=1}^{M}\left(\frac{1}{\sqrt{2\pi\delta}}\right)^se^{-\left(\frac{\bm{\xi}^{(k)}}{\sqrt{\delta}}\right)^2},
\ee
\noindent and use stochastic gradient descent to find the minimum. In the specific case we considered in our simulations, the noise simply affected the ``susceptible'' and ``exposed'' compartments, where the usual infection rate is multiplied by a factor $(1+\xi)$, thus modelling a stochastic fluctuation of the infection rate. This is arguably the most interesting term to study stochastic fluctuations, since the product $SI$ appearing in the differential equation is at the origin of the nonlinearity of the SEIR model. 

In the next section, we will use the same idea to carry out policy optimizations in scenarios where the disease's evolution is influenced by a finite number of discrete random variables.

\section{Optimization over discrete policies of disease control}
\label{discrete_app}
The section above explains how to conduct optimizations over disease control policies, as long as the parameters $\bm{\mu}$ defining the policy are allowed to vary all over $\R^n$. Some policies, though, are by their very nature, discrete. For instance, on day $t$ we can either declare a lockdown ($s(t)=1$) or not declare a lockdown ($s(t)=0$). As we discuss in section \ref{applications} in the main text, the effect of a lockdown policy in the evolution of the disease can be modeled by introducing a term in eq. (\ref{original}) that is proportional to $s(t)$. Namely, $G^i(t,\bm{x};\bm{\mu})=\hat{G}^i(t,\bm{x};\bm{\mu})+s(t)F^i(\bm{x})$. 

To optimize over such discrete policies via gradient descent methods, one could think of introducing a continuous variable $\lambda\in \R$ and writing its effect on the disease's equations of motion by means of a piece-wise continuous function of $\lambda$, e.g.: $s(t) =\Theta(\lambda)$, for $t\in[t_1,t_2]$. Here $\Theta(z)$ denotes the Heaviside function (i.e., $\Theta(z)$ equals $1$ for $z\geq 0$, or $0$, otherwise). In this case, however, the gradient method would not work, since the Heaviside function has zero derivative everywhere except at $0$. In every iteration of Adam, $\nabla_\lambda A$ would be null, and so $\lambda^{(k)}=\lambda^{(0)}$ for all $k$.

Applying the gradient method to optimize over policies for disease control involving discrete government interventions is therefore not straightforward. In the following, we propose two heuristics to tackle this problem.

\subsection{Optimization over deterministic discrete policies through non-deterministic discrete policies}

Suppose, for the time being, that our lockdown policy were probabilistic, i.e., at each week $k$, we declare a lockdown with probability $p_k(1)=\sigma(\tilde{s}_k)$; otherwise, with probability $p_k(0)=1-\sigma(\tilde{s}_k)$, we let the population roam freely. We wish to minimize our \emph{average} objective function, that is, the expression

\be
\bar{A}(\tilde{\bm{s}}, \bm{\mu})=\sum_{c_1,...,c_q=0,1} \prod_{k=1}^qp_k(c_k)\int dt {\cal L}\left(t, \bm{\mu}, \bm{x}(t;c,\bm{\mu}, \bm{x}_0)\right),
\label{proba_obj}
\ee 
\noindent where $c$ is the whole vector of weekly lockdowns, and $\bm{\mu}$ corresponds to the continuous parameters of the policy, e.g.: vaccination rates.

In principle, we could apply gradient descent to minimize (\ref{proba_obj}). Estimating the exact gradient of the above expression is, however, unrealistic, as it involves summing a number of terms exponential in the number of weeks $q$. Instead, we will produce a random unbiased estimate of the gradient and invoke stochastic gradient descent methods, see Section \ref{gradient_sec}. 

Let us first differentiate Eq. (\ref{proba_obj}) with respect to the continuous variables $\bm{\mu}$. The result is
\be\label{eq:avg_dSdmu}
\frac{\partial\bar{A}(\tilde{\bm{s}}, \bm{\mu})}{\partial\mu_j}=\sum_{c_1,...,c_q=0,1} \prod_{k=1}^qp_k(c_k)v_j(\tilde{\bm{s}}, \bm{\mu}|c)=\langle v_j(\tilde{\bm{s}}, \bm{\mu}|c)\rangle_c,
\ee 
\noindent where
\be
v_j(\tilde{\bm{s}}, \bm{\mu}|c)=\int dt \frac{\partial{\cal L}\left(t, \bm{\mu}, \bm{x}(t;c,\bm{\mu}, \bm{x}_0)\right)}{\partial \mu_j}
\label{Ds}
\ee 
 and the components of the random variable $c\in\{0,1\}^q$ are generated by sequentially sampling from the Bernouilli distributions $(1-\sigma(\tilde{s}_k), \sigma(\tilde{s}_k))_k$. Note that the expression in the integrand of (\ref{Ds}) can be computed using the techniques discussed in Section \ref{optimization_app}.

Differentiating Eq. (\ref{proba_obj}) with respect to $\tilde{s}_k$ we find that

\be\label{eq:avg_dSds}
\frac{\partial\bar{A}(\tilde{\bm{s}}, \bm{\mu})}{\partial \tilde{s}_k}=\sum_{a=0,1} \left\langle w_k(\tilde{\bm{s}}, \bm{\mu}|c^{(k,a)})\right\rangle_{c^{(k,a)}},
\ee
\noindent with
\be
w_k(\tilde{\bm{s}}, \bm{\mu}|c^{(k,a)})=\frac{\partial p_k(a)}{\partial \tilde{s}_k}\int dt {\cal L}\left(t, \bm{\mu}, \bm{x}(t; c^{(k,a)},\bm{\mu}, \bm{x}_0)\right), \text{ for } a=0,1\,
\ee
and the average $\left\langle w_k(\tilde{\bm{s}}, \bm{\mu}|c^{(k,a)})\right\rangle_{c^{(k,a)}}$ is obtained via sampling over the product of Bernoulli distributions for $c^{(k,a)}_1,\ldots,c^{(k,a)}_{k-1},c^{(k,a)}_{k+1},\ldots,c^{(k,a)}_q$ and fixing $c^{(k,a)}_k = a$.

Putting all this together, we have that the random vectors $v, w$ satisfy

\begin{align}
&\left\langle v\right\rangle = \nabla_{\bm{\mu}}\bar{A}(\tilde{\bm{s}}, \bm{\mu})\nonumber\\
&\left\langle w\right\rangle = \nabla_{s}\bar{A}(\tilde{\bm{s}}, \bm{\mu}).
\end{align}
\noindent Since both vectors can be sampled efficiently, we can use them (and their averages) to optimize over $\bar{A}(\tilde{\bm{s}}, \bm{\mu})$ via stochastic gradient descent.

At this point, the reader might object that our original goal was to minimize (\ref{functional1}) over policies with \emph{deterministic} lockdown. Very conveniently, independently of the initial values of $s, \bm{\mu}$, the stochastic gradient method will converge to a policy $p^\star, \bm{\mu}^\star$ such that the deterministic policy with the same continuous parameters $\bm{\mu}^\star$ and lockdown given by
\begin{align}
c^\star_k = 
\begin{cases}
0 & \text{for } p^\star_k(0)>1/2\\
1 & \text{otherwise} 
\end{cases} 
\label{determine}
\end{align}
\noindent has the same objective value.
Indeed, for $k\in\{1,...,n\}$, fix $\{s^\star_j:j\not=k\}$,  then

\be
\bar{A}(\tilde{\bm{s}}^\star, \bm{\mu}^\star) =\sum_{a=0,1} p^\star_k(a) A(\tilde{\bm{s}}^\star,\bm{\mu}^\star|a_k =a),
\ee
\noindent with 

\begin{align}
&A(\tilde{\bm{s}}^\star,\bm{\mu}^\star|a_k = a)\equiv\nonumber\\
&\sum_{c_1,...c_{k-1},c_{k+1},...,c_q=0,1} \prod_{j\not=k}^qp^\star_j(c_j)\int dt {\cal L}\left(t,\bm{\mu}^\star, c, \bm{x}(t;c_1,...,c_{k-1}, a,c_{k+1},...,c_q,\bm{\mu}^\star, \bm{x}_0)\right).
\end{align}
\noindent Since $p^\star,\bm{\mu}^\star$ is a local minimum of $\bar{A}(\tilde{\bm{s}}^\star, \bm{\mu}^\star)$, it follows that, either  

\be
\bar{A}(\tilde{\bm{s}}^\star, \bm{\mu}^\star)=A(\tilde{\bm{s}}^\star,\bm{\mu}^\star|a_k = 0)=A(\tilde{\bm{s}}^\star,\bm{\mu}^\star|a_k = 1),
\ee
\noindent or, for some $a\in\{0,1\}$, 

\be
\bar{A}(\tilde{\bm{s}}^\star, \bm{\mu}^\star)=A(\tilde{\bm{s}}^\star,\bm{\mu}^\star|a_k = a)<A(\tilde{\bm{s}}^\star,\bm{\mu}^\star|a_k = 1-a),
\ee
\noindent with $p^\star_k(a)=1, p^\star_k(1-a)=0$. In either case, fixing $c_k$ through the procedure (\ref{determine}) cannot increase the average value of the objective function. Iterating over $k=1,...,q$, we prove the claim.

\subsubsection{Generalization to optimizations over adaptive policies}

The method described above can be easily extended to tackle optimization problems over weekly discrete adaptive policies. Consider an adaptive policy where the government intervention $c_k\in\{0,1\}$ on week $k$ is decided on week $k-l$ by sampling from a Bernoulli distribution dependent on the value $\bm{\theta}^k$ of the cell state on week $k-l$ (remember from Appendix \ref{optimization_app} that the cell state at time $t$ represents the government's internal memory and is given by some of the entries of $\bm{x}(t)$). The functional form of this distribution is determined by a vector of parameters $\bm{\nu}^k$. We thus have that $p_k(c_k)=p_k(c_k|\bm{\theta}^k;\bm{\nu}^k)$. 

Note that $\bm{x}^k=\bm{x}^k(c_1,...,c_{k-l-1})$. Starting from the initial conditions $\bm{x}_0$, the average objective function can therefore be estimated by propagating $\bm{x}$ week by week, computing the contribution to (\ref{functional1}) and sampling each $c_k$ at time $7(k-l)$ as we go along. If we instead compute the contribution to the gradient of (\ref{functional1}) with respect to the continuous variables $\bm{\mu}$, we will have a statistical estimate of the gradient of the average objective function (with respect to $\bm{\mu}$).

Estimating the gradient of the average objective function with respect to the variables $\bm{\nu}^k$ is done similarly, by averaging over the appropriate Markov chain. More specifically,

\be\label{eq:avg_dSds_adapt}
\nabla_{\bm{\nu}^k}\bar{A}(\tilde{\bm{s}}, \bm{\mu})=\sum_{a=0,1} \left\langle w_k(\tilde{\bm{\nu}}, \bm{\mu}|c^{(k,a)})\right\rangle_{c^{(k,a)}},
\ee
\noindent with
\be
w_k(\tilde{\bm{\nu}}, \bm{\mu}|c^{(k,a)})=\nabla_{\bm{\nu}^k}p_k(a|\bm{\theta}^k;\bm{\nu}^k)\times \int dt {\cal L}\left(t, \bm{\mu}, \bm{x}(t; c^{(k,a)},\bm{\mu}, \bm{x}_0)\right), \text{ for } a=0,1.
\ee
\noindent This time, $c^{(k,a)}$ is sampled sequentially while we solve the differential equation, as described before to estimate the average of the objective function. The only difference is that, at time $t_0+7(k-l)$, instead of sampling $c_k$, we set it equal to $a$.

Now, let us assume that, for $\bm{y}_1\not=\bm{y}_2$, the family of functions $p_k(a|\bm{\theta}^k; \bm{\nu}^k)$ available is rich enough to regard $p_k(a|\bm{\theta}^k=\bm{y}_1; \bm{\nu}^k), p_k(a|\bm{\theta}^k=\bm{y}_2; \bm{\nu}^k)$ as independent. Then one can argue as for non-adaptive policies and conclude that the gradient method will converge to a deterministic policy.

\subsection{Optimization over discrete policies with continuous lockdown times}

\begin{figure}
  \centering
  \includegraphics[width=14 cm]{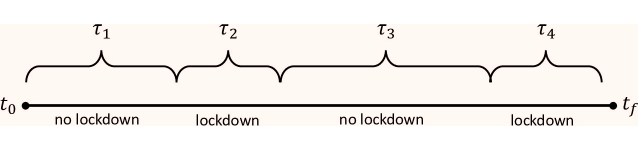}
  \caption{\textbf{Lockdowns (discrete policies) parametrized with continuous time.} In this example there are two lockdowns and two periods of freedom. The algorithm finds the optimal distribution and minimises the total time $\tau_2 + \tau_4$ in lockdown.}
  \label{continuous_param_fig}
\end{figure}

Our second heuristic to devise discrete policies for disease control requires considering lockdown policies of the following form:

\begin{align}
\label{smooth_conf}
s(t) =
\begin{cases}
0 & \text{ if } (t-t_0)\leq \tau_1 \text{ or } \sum_{k=1}^{2i}\tau_k\leq (t-t_0)\leq \sum_{k=1}^{2i+1}\tau_k, \text{ for some } i,\\
1 &\text{otherwise}. 
\end{cases} 
\end{align}
Here $t_0$ is fixed and the variables $\{\tau_k\}_{k=1}^{n}$ are assumed to be non-negative and to add up to $t_f-t_0$; this policy can hence be parametrized by a vector $\bm{\mu}\in\R^n$, with $\bm{\tau} =(t_f-t_0)\mbox{softmax}(\bm{\mu})$, where $\mbox{softmax}(\bm{\mu})$ denotes the vector $\bm{\nu}$ with components $\nu_i=\frac{\mbox{exp}(\mu_i)}{\sum_j\mbox{exp}(\mu_j)}$. Intuitively, $\{\tau_k\}_{i=1}^{n}$ divide the interval $[t_0, t_f]$ into $n$ different parts. In each part, lockdown is alternatively declared ($s=1$) or suspended ($s=0$), see Figure \ref{continuous_param_fig}.

\begin{figure}
  \centering
  \includegraphics[width=10 cm]{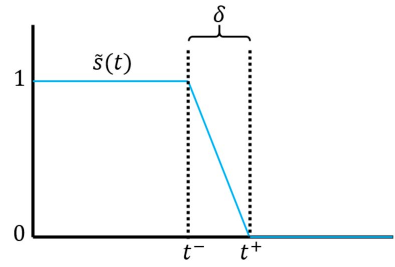}
  \caption{\textbf{Modified continuous function $\tilde{s}(t)$.}}
  \label{s_tilde_fig}
\end{figure}

At time $t$, the disease's basic reproduction number is given by (\ref{effect_trans}), with $s(t)$ defined as above. To find out $y^{i}_j\equiv\frac{\partial x^i(t;\bm{\mu})}{\partial \bm{\mu}_j}$, we invoke Eq. (\ref{extra}). In computing the term 

\be
\frac{\partial G^i}{\partial \mu_j}=\sum_{k}\frac{\partial G^i}{\partial \tau_k}\frac{\partial \tau_k}{\partial \mu_j}, 
\ee
\noindent we have the problem that, due to (\ref{smooth_conf}), $G$ is not continuous or differentiable. To work our way out, we approximate $s(t)$ by a piece-wise continuous function with bounded derivative that transitions from $0$ to $1$ (or viceversa) linearly and in time $\delta\ll 1$, see Figure \ref{s_tilde_fig}; later we will take the limit $\delta\to 0$. 

The new function $\tilde{s}(t)$ has zero derivative with respect to $\mu_i$, except for $t$ satisfying

\be
t^-:= (t_0 + \sum_{k=1}^{u}\tau_k) 
\leq t\leq 
(t_0 +\sum_{k=1}^{u}\tau_k+\delta )
=:t^+.
\ee
In that case, the derivative of $\hat{s}$ with respect to $\tau_j$, with $j\leq u$, will (approximately) be 

\begin{align}
\frac{\partial}{\partial \tau_j}\hat{s}(t) =&\frac{(-1)^u}{\delta}. 
\end{align}
\noindent The derivative with respect to any of the variables $\{\tau_j:j>u\}$ is zero.
The dominant term on the right-hand side of (\ref{extra}) for $t\in[t^-,t^+]$ is therefore 

\begin{align}
&\frac{\partial G^i}{\partial \mu_j} = \frac{\partial G^i(t^-,\bm{x}(t^-))}{\partial \tilde{s}}\sum_{k=1}^u\frac{\partial \tilde{s}}{\partial \tau_k}\frac{\partial \tau_k}{\partial \mu_j}\nonumber\\
&=(t_f-t_0)\frac{(-1)^u}{\delta}\frac{\partial G^i(t^-,\bm{x}(t^-))}{\partial \tilde{s}}\sum_{k=1}^u\frac{\partial \mbox{softmax}(\bm{\mu})_k}{\partial \mu_j}=:\frac{K^i_j(t^-, \bm{x}(t^-),\bm{\mu}, u)}{\delta}.
\end{align}
\noindent Since the evolution takes place for time $\delta=t^+-t^-$, we have that $y^i_j(t^+)\approx y^i_j(t^-) + K^i_j(t^-, \bm{x}(t^-),\bm{\mu}, m)$. Taking the limit $\delta\to 0$, we have that the evolution of $y^{i}_j$ is determined by the following prescription:

\begin{enumerate}
\item
$y^{i}_j(t_0)=0$.
\item
Let $t=t_0 + \sum_{1=0}^u \tau_k$, for some $u$. Then $y^{i}_j$ is updated by the rule

\be
y^{i}_j(t)\to y^{i}_j(t) + K^i_j(t, \bm{x},\bm{\mu}, u).
\ee
\item
For all other values of $t$, $y^{i}_j$ continuously evolves via the equation

\be
\frac{dy^i_j}{dt}=\sum_l \frac{\partial G^i(t, \bm{x};\bm{\mu})}{\partial x^l}y^l_j; i=1,...,m, j =1,...,n.
\ee

\end{enumerate}

\section{Continuous-space population models}
\label{spatial_app}
Even though the focus of this article is that of compartmental models of the form (\ref{original}), one can also apply the principles of gradient descent for policy optimizations on dynamical systems governed by partial differential equations. Consider, e.g., the scenario studied in \cite{raccoons}, where the authors model the spread of rabies in raccoons across a realistic landscape $\Omega\subset \R^2$ through a system of reaction-diffusion equations of the form

\be
\frac{\partial}{\partial t} \bm{u} -\mbox{div}(\nu \nabla \bm{u})=A(\bm{u})\bm{u}.
\label{diffusion}
\ee
\noindent In this equation, the three entries of the vector field $u(t,X,Y)\in\R^3$ respectively denote the number of susceptible, exposed and infected individuals at time $t$ in position $X, Y$. $\nu, A$ are $3\times 3$ matrices that, in principle, might depend on some controllable parameters $\bm{\mu}$. This equation is to be solved under the initial conditions $u(0, X, Y)=u_0(X, Y)$ and the homogeneous von Neumann boundary conditions

\be
\nabla \bm{u}(t, X, Y)\cdot \bm{n}(X,Y) =0, \mbox{ for } (X,Y)\in \partial\Omega,
\label{vonNeumann}
\ee
\noindent where $\bm{n}(X,Y)\in\R^2$ denotes the vector normal to the contour $\partial\Omega$ at location $(X,Y)$. The authors of \cite{raccoons} solve this equation numerically via the Finite Element Method (FEM) \cite{fem}.

Suppose that we wished to optimize the policy parameters $\bm{\mu}\in\R^n$ over some functional $A$ depending on $\bm{u}(t, X, Y;\bm{\mu}, u_0)$ (instead of $\bm{x}(t;\bm{\mu}, \bm{x}_0)$) via the gradient method. Then at some point we would need to compute the quantities $v^i_j(t, X, Y;\bm{\mu},u_0)\equiv \frac{\partial u^i(t, X, Y;\bm{\mu},u_0)}{\partial\mu_j}$. Let $\bm{v_j}\in\R^3$ be the vector with components $v^i_j$ and differentiate both (\ref{diffusion}) and (\ref{vonNeumann}) with respect to $\mu_i$. This results in the equation

\begin{align}
&\frac{\partial}{\partial t} \bm{v_j} -\mbox{div}(\frac{\partial\nu}{\partial \mu_j} \nabla \bm{u} + \nu\nabla \bm{v_j})=(\frac{\partial A}{\partial \mu_j} + \frac{\partial A}{\partial u_i}v^i_j)\bm{u} + A(\bm{u})\bm{v_j}\nonumber\\
&\nabla \bm{v_j}(t, X, Y)\cdot \bm{n} =0, \mbox{ for } (X,Y)\in \partial\Omega.
\label{diffusion_deriv}
\end{align}
\noindent Since $\bm{u}(0, X, Y;\bm{\mu}, \bm{u_0})$ does not depend on $\bm{\mu}$, this new diffusion equation must be solved for the initial conditions $\bm{v_j}(X, Y, 0)=0$. This can be achieved numerically in the same way that the authors of \cite{raccoons} solved Eq. (\ref{diffusion}), that is, via the FEM.

\end{appendix}

\end{document}